\documentclass[conference,a4paper]{IEEEtran}
\IEEEoverridecommandlockouts
\usepackage[acronyms,nonumberlist,nopostdot,nomain,nogroupskip,acronymlists={hidden}]{glossaries}
\newglossary[algh]{hidden}{acrh}{acnh}{Hidden Acronyms}

\usepackage{cite}
\usepackage{amsmath,amssymb,amsfonts}
\usepackage{algorithmic}
\usepackage{textcomp}
\usepackage{xcolor}
\usepackage{balance}
\usepackage{booktabs}
\usepackage{makecell}
\usepackage{multirow} % Required for merging rows

\usepackage{tikz}
\usepackage{pgfplots}
\pgfplotsset{compat=newest}
\pgfplotsset{plot coordinates/math parser=false}
\newlength\fheight
\newlength\fwidth
\usetikzlibrary{plotmarks,patterns,decorations.pathreplacing,backgrounds,calc,arrows,arrows.meta,spy,matrix,scopes}
\usepgfplotslibrary{patchplots,groupplots}
\usepackage{tikzscale}
\pgfplotsset{compat=1.18}

\def\BibTeX{{\rm B\kern-.05em{\sc i\kern-.025em b}\kern-.08em
    T\kern-.1667em\lower.7ex\hbox{E}\kern-.125emX}}

\expandafter\def\csname ver@fixltx2e.sty\endcsname{}

\newif\ifexttikz
\exttikzfalse
\ifexttikz
	\usetikzlibrary{external}
\fi

\usepackage[font=footnotesize]{caption}
\usepackage{subcaption}
\usepackage{soul}

\newacronym{3gpp}{3GPP}{3rd Generation Partnership Project}
\newacronym{4g}{4G}{4th generation}
\newacronym{5g}{5G}{5th generation}
\newacronym{6g}{6G}{6th generation}
\newacronym{5gc}{5GC}{5G Core}
\newacronym{adc}{ADC}{Analog to Digital Converter}
\newacronym{aerpaw}{AERPAW}{Aerial Experimentation and Research Platform for Advanced Wireless}
\newacronym{ai}{AI}{Artificial Intelligence}
\newacronym{aimd}{AIMD}{Additive Increase Multiplicative Decrease}
\newacronym{am}{AM}{Acknowledged Mode}
\newacronym{amc}{AMC}{Adaptive Modulation and Coding}
\newacronym{amf}{AMF}{Access and Mobility Management Function}
\newacronym{aops}{AOPS}{Adaptive Order Prediction Scheduling}
\newacronym{api}{API}{Application Programming Interface}
\newacronym{apn}{APN}{Access Point Name}
\newacronym{ap}{AP}{Application Protocol}
\newacronym{aqm}{AQM}{Active Queue Management}
\newacronym{ausf}{AUSF}{Authentication Server Function}
\newacronym{avc}{AVC}{Advanced Video Coding}
\newacronym{awgn}{AGWN}{Additive White Gaussian Noise}
\newacronym{balia}{BALIA}{Balanced Link Adaptation Algorithm}
\newacronym{bbu}{BBU}{Base Band Unit}
\newacronym{bdp}{BDP}{Bandwidth-Delay Product}
\newacronym{ber}{BER}{Bit Error Rate}
\newacronym{bf}{BF}{Beamforming}
\newacronym{bler}{BLER}{Block Error Rate}
\newacronym{brr}{BRR}{Bayesian Ridge Regressor}
\newacronym{bs}{BS}{Base Station}
\newacronym{bsr}{BSR}{Buffer Status Report}
\newacronym{bss}{BSS}{Business Support System}
\newacronym{ca}{CA}{Carrier Aggregation}
\newacronym{caas}{CaaS}{Connectivity-as-a-Service}
\newacronym{cb}{CB}{Code Block}
\newacronym{cc}{CC}{Congestion Control}
\newacronym{ccid}{CCID}{Congestion Control ID}
\newacronym{cco}{CC}{Carrier Component}
\newacronym{cdd}{CDD}{Cyclic Delay Diversity}
\newacronym{cdf}{CDF}{Cumulative Distribution Function}
\newacronym{cdn}{CDN}{Content Distribution Network}
\newacronym{cir}{CIR}{Channel Impulse Response}
\newacronym{cli}{CLI}{Command-line Interface}
\newacronym{cn}{CN}{Core Network}
\newacronym{cnn}{CNN}{Convolutional Neural Network}
\newacronym{codel}{CoDel}{Controlled Delay Management}
\newacronym{comac}{COMAC}{Converged Multi-Access and Core}
\newacronym{cord}{CORD}{Central Office Re-architected as a Datacenter}
\newacronym{cornet}{CORNET}{COgnitive Radio NETwork}
\newacronym{cosmos}{COSMOS}{Cloud Enhanced Open Software Defined Mobile Wireless Testbed for City-Scale Deployment}
\newacronym{cots}{COTS}{Commercial Off-the-Shelf}
\newacronym{cp}{CP}{Control Plane}
\newacronym{cyp}{CP}{Cyclic Prefix}
\newacronym{up}{UP}{User Plane}
\newacronym{cpu}{CPU}{Central Processing Unit}
\newacronym{cqi}{CQI}{Channel Quality Information}
\newacronym{cql}{CQL}{Conservative Q-Learning}
\newacronym{cr}{CR}{Cognitive Radio}
\newacronym{cran}{CRAN}{Cloud \gls{ran}}
\newacronym{crs}{CRS}{Cell Reference Signal}
\newacronym{csi}{CSI}{Channel State Information}
\newacronym{csirs}{CSI-RS}{Channel State Information - Reference Signal}
\newacronym{cu}{CU}{Central Unit}
\newacronym{cucp}{CU-CP}{Central Unit Control Plane}
\newacronym{cuup}{CU-UP}{Central Unit User Plane}
\newacronym{d2tcp}{D$^2$TCP}{Deadline-aware Data center TCP}
\newacronym{d3}{D$^3$}{Deadline-Driven Delivery}
\newacronym{dac}{DAC}{Digital to Analog Converter}
\newacronym{dag}{DAG}{Directed Acyclic Graph}
\newacronym{das}{DAS}{Distributed Antenna System}
\newacronym{dash}{DASH}{Dynamic Adaptive Streaming over HTTP}
\newacronym{dc}{DC}{Dual Connectivity}
\newacronym{dccp}{DCCP}{Datagram Congestion Control Protocol}
\newacronym{dce}{DCE}{Direct Code Execution}
\newacronym{dci}{DCI}{Downlink Control Information}
\newacronym{dctcp}{DCTCP}{Data Center TCP}
\newacronym{dl}{DL}{Downlink}
\newacronym{dmr}{DMR}{Deadline Miss Ratio}
\newacronym{dmrs}{DMRS}{DeModulation Reference Signal}
\newacronym{dqn}{DQN}{Deep Q-Network}
\newacronym{drlcc}{DRL-CC}{Deep Reinforcement Learning Congestion Control}
\newacronym{drs}{DRS}{Discovery Reference Signal}
\newacronym{du}{DU}{Distributed Unit}
\newacronym{ee}{EE}{Energy Efficiency}
\newacronym{e2e}{E2E}{end-to-end}
\newacronym{earfcn}{EARFCN}{E-UTRA Absolute Radio Frequency Channel Number}
\newacronym{ecaas}{ECaaS}{Edge-Cloud-as-a-Service}
\newacronym{ecn}{ECN}{Explicit Congestion Notification}
\newacronym{edf}{EDF}{Earliest Deadline First}
\newacronym{embb}{eMBB}{Enhanced Mobile Broadband}
\newacronym{empower}{EMPOWER}{EMpowering transatlantic PlatfOrms for advanced WirEless Research}
\newacronym{enb}{eNB}{evolved Node Base}
\newacronym{endc}{EN-DC}{E-UTRAN-\gls{nr} \gls{dc}}
\newacronym{epc}{EPC}{Evolved Packet Core}
\newacronym{eps}{EPS}{Evolved Packet System}
\newacronym{es}{ES}{Edge Server}
\newacronym{etsi}{ETSI}{European Telecommunications Standards Institute}
\newacronym[firstplural=Estimated Times of Arrival (ETAs)]{eta}{ETA}{Estimated Time of Arrival}
\newacronym{eutran}{E-UTRAN}{Evolved Universal Terrestrial Access Network}
\newacronym{faas}{FaaS}{Function-as-a-Service}
\newacronym{fapi}{FAPI}{Functional Application Platform Interface}
\newacronym{fdd}{FDD}{Frequency Division Duplexing}
\newacronym{fdm}{FDM}{Frequency Division Multiplexing}
\newacronym{fdma}{FDMA}{Frequency Division Multiple Access}
\newacronym{fed4fire}{FED4FIRE+}{Federation 4 Future Internet Research and Experimentation Plus}
\newacronym{fir}{FIR}{Finite Impulse Response}
\newacronym{fit}{FIT}{Future \acrlong{iot}}
\newacronym{fpga}{FPGA}{Field Programmable Gate Array}
\newacronym{fr1}{FR1}{Frequency Range 1}
\newacronym{fr2}{FR2}{Frequency Range 2}
\newacronym{fs}{FS}{Fast Switching}
\newacronym{fscc}{FSCC}{Flow Sharing Congestion Control}
\newacronym{ftp}{FTP}{File Transfer Protocol}
\newacronym{fw}{FW}{Flow Window}
\newacronym{ge}{GE}{Gaussian Elimination}
\newacronym{gnb}{gNB}{Next Generation Node Base}
\newacronym{gop}{GOP}{Group of Pictures}
\newacronym{gpr}{GPR}{Gaussian Process Regressor}
\newacronym{gpu}{GPU}{Graphics Processing Unit}
\newacronym{gtp}{GTP}{GPRS Tunneling Protocol}
\newacronym{gtpc}{GTP-C}{GPRS Tunnelling Protocol Control Plane}
\newacronym{gtpu}{GTP-U}{GPRS Tunnelling Protocol User Plane}
\newacronym{gtpv2c}{GTPv2-C}{\gls{gtp} v2 - Control}
\newacronym{gw}{GW}{Gateway}
\newacronym{harq}{HARQ}{Hybrid Automatic Repeat reQuest}
\newacronym{hetnet}{HetNet}{Heterogeneous Network}
\newacronym{hh}{HH}{Hard Handover}
\newacronym{hol}{HOL}{Head-of-Line}
\newacronym{hqf}{HQF}{Highest-quality-first}
\newacronym{hss}{HSS}{Home Subscription Server}
\newacronym{http}{HTTP}{HyperText Transfer Protocol}
\newacronym{ia}{IA}{Initial Access}
\newacronym{iab}{IAB}{Integrated Access and Backhaul}
\newacronym{ic}{IC}{Incident Command}
\newacronym{ietf}{IETF}{Internet Engineering Task Force}
\newacronym{imsi}{IMSI}{International Mobile Subscriber Identity}
\newacronym{imt}{IMT}{International Mobile Telecommunication}
\newacronym{iot}{IoT}{Internet of Things}
\newacronym{ip}{IP}{Internet Protocol}
\newacronym{itu}{ITU}{International Telecommunication Union}
\newacronym{kpi}{KPI}{Key Performance Indicator}
\newacronym{kpm}{KPM}{Key Performance Measurement}
\newacronym{kvm}{KVM}{Kernel-based Virtual Machine}
\newacronym{los}{LoS}{Line-of-Sight}
\newacronym{lsm}{LSM}{Link-to-System Mapping}
\newacronym{lstm}{LSTM}{Long Short Term Memory}
\newacronym{lte}{LTE}{Long Term Evolution}
\newacronym{lxc}{LXC}{Linux Container}
\newacronym{m2m}{M2M}{Machine to Machine}
\newacronym{mac}{MAC}{Medium Access Control}
\newacronym{manet}{MANET}{Mobile Ad Hoc Network}
\newacronym{mano}{MANO}{Management and Orchestration}
\newacronym{mc}{MC}{Multi-Connectivity}
\newacronym{mcc}{MCC}{Mobile Cloud Computing}
\newacronym{mchem}{MCHEM}{Massive Channel Emulator}
\newacronym{mcs}{MCS}{Modulation and Coding Scheme}
\newacronym{mec}{MEC}{Multi-access Edge Computing}
\newacronym{mec2}{MEC}{Mobile Edge Cloud}
\newacronym{mfc}{MFC}{Mobile Fog Computing}
\newacronym{mgen}{MGEN}{Multi-Generator}
\newacronym{mi}{MI}{Mutual Information}
\newacronym{mib}{MIB}{Master Information Block}
\newacronym{miesm}{MIESM}{Mutual Information Based Effective SINR}
\newacronym{mimo}{MIMO}{Multiple Input, Multiple Output}
\newacronym{ml}{ML}{Machine Learning}
\newacronym{mlr}{MLR}{Maximum-local-rate}
\newacronym[plural=\gls{mme}s,firstplural=Mobility Management Entities (MMEs)]{mme}{MME}{Mobility Management Entity}
\newacronym{mmtc}{mMTC}{Massive Machine-Type Communications}
\newacronym{mmwave}{mmWave}{millimeter wave}
\newacronym{mpdccp}{MP-DCCP}{Multipath Datagram Congestion Control Protocol}
\newacronym{mptcp}{MPTCP}{Multipath TCP}
\newacronym{mr}{MR}{Maximum Rate}
\newacronym{mrdc}{MR-DC}{Multi \gls{rat} \gls{dc}}
\newacronym{mse}{MSE}{Mean Square Error}
\newacronym{mss}{MSS}{Maximum Segment Size}
\newacronym{mt}{MT}{Mobile Termination}
\newacronym{mtd}{MTD}{Machine-Type Device}
\newacronym{mtu}{MTU}{Maximum Transmission Unit}
\newacronym{mumimo}{MU-MIMO}{Multi-user \gls{mimo}}
\newacronym{mvno}{MVNO}{Mobile Virtual Network Operator}
\newacronym{nalu}{NALU}{Network Abstraction Layer Unit}
\newacronym{nas}{NAS}{Network Attached Storage}
\newacronym{nat}{NAT}{Network Address Translation}
\newacronym{nbiot}{NB-IoT}{Narrow Band IoT}
\newacronym{nfv}{NFV}{Network Function Virtualization}
\newacronym{nfvi}{NFVI}{Network Function Virtualization Infrastructure}
\newacronym{ni}{NI}{Network Interfaces}
\newacronym{nic}{NIC}{Network Interface Card}
\newacronym{nlos}{NLoS}{Non-Line-of-Sight}
\newacronym{now}{NOW}{Non Overlapping Window}
\newacronym{nsm}{NSM}{Network Service Mesh}
\newacronym[type=hidden]{nr}{NR}{New Radio}
\newacronym{nextg}{NextG}{Next Generation}
\newacronym{nrf}{NRF}{Network Repository Function}
\newacronym{nsa}{NSA}{Non Stand Alone}
\newacronym{nse}{NSE}{Network Slicing Engine}
\newacronym{nssf}{NSSF}{Network Slice Selection Function}
\newacronym{o2i}{O2I}{Outdoor to Indoor}
\newacronym{oai}{OAI}{OpenAirInterface}
\newacronym{oaicn}{OAI-CN}{\gls{oai} \acrlong{cn}}
\newacronym{oairan}{OAI-RAN}{\acrlong{oai} \acrlong{ran}}
\newacronym{oam}{OAM}{Operations, Administration and Maintenance}
\newacronym{ofdm}{OFDM}{Orthogonal Frequency Division Multiplexing}
\newacronym{olia}{OLIA}{Opportunistic Linked Increase Algorithm}
\newacronym{omec}{OMEC}{Open Mobile Evolved Core}
\newacronym{onap}{ONAP}{Open Network Automation Platform}
\newacronym{onf}{ONF}{Open Networking Foundation}
\newacronym{onos}{ONOS}{Open Networking Operating System}
\newacronym{oom}{OOM}{\gls{onap} Operations Manager}
\newacronym{opnfv}{OPNFV}{Open Platform for \gls{nfv}}
\newacronym[type=hidden]{oran}{O-RAN}{Open \gls{ran}}
\newacronym{orbit}{ORBIT}{Open-Access Research Testbed for Next-Generation Wireless Networks}
\newacronym{os}{OS}{Operating System}
\newacronym{osm2}{OSM}{Open Street Map}
\newacronym{oss}{OSS}{Operations Support System}
\newacronym{pa}{PA}{Position-aware}
\newacronym{pase}{PASE}{Prioritization, Arbitration, and Self-adjusting Endpoints}
\newacronym{pawr}{PAWR}{Platforms for Advanced Wireless Research}
\newacronym{pbch}{PBCH}{Physical Broadcast Channel}
\newacronym{pcef}{PCEF}{Policy and Charging Enforcement Function}
\newacronym{pcfich}{PCFICH}{Physical Control Format Indicator Channel}
\newacronym{pcrf}{PCRF}{Policy and Charging Rules Function}
\newacronym{pdcch}{PDCCH}{Physical Downlink Control Channel}
\newacronym{pdcp}{PDCP}{Packet Data Convergence Protocol}
\newacronym{pdsch}{PDSCH}{Physical Downlink Shared Channel}
\newacronym{pdu}{PDU}{Packet Data Unit}
\newacronym{pf}{PF}{Proportional Fair}
\newacronym{pgw}{PGW}{Packet Gateway}
\newacronym{phich}{PHICH}{Physical Hybrid ARQ Indicator Channel}
\newacronym{phy}{PHY}{Physical}
\newacronym{pl}{PL}{Path Loss}
\newacronym{pmch}{PMCH}{Physical Multicast Channel}
\newacronym{pmi}{PMI}{Precoding Matrix Indicators}
\newacronym{powder}{POWDER}{Platform for Open Wireless Data-driven Experimental Research}
\newacronym{ppo}{PPO}{Proximal Policy Optimization}
\newacronym{ppp}{PPP}{Poisson Point Process}
\newacronym{prach}{PRACH}{Physical Random Access Channel}
\newacronym{prb}{PRB}{Physical Resource Block}
\newacronym{psnr}{PSNR}{Peak Signal to Noise Ratio}
\newacronym{pss}{PSS}{Primary Synchronization Signal}
\newacronym{pucch}{PUCCH}{Physical Uplink Control Channel}
\newacronym{pusch}{PUSCH}{Physical Uplink Shared Channel}
\newacronym{qam}{QAM}{Quadrature Amplitude Modulation}
\newacronym{qci}{QCI}{\gls{qos} Class Identifier}
\newacronym{qoe}{QoE}{Quality of Experience}
\newacronym{qos}{QoS}{Quality of Service}
\newacronym{quic}{QUIC}{Quick UDP Internet Connections}
\newacronym{rach}{RACH}{Random Access Channel}
\newacronym{ran}{RAN}{Radio Access Network}
\newacronym[firstplural=Radio Access Technologies (RATs)]{rat}{RAT}{Radio Access Technology}
\newacronym{rbg}{RBG}{Resource Block Group}
\newacronym{rcn}{RCN}{Research Coordination Network}
\newacronym{rc}{RC}{RAN Control}
\newacronym{rec}{REC}{Radio Edge Cloud}
\newacronym{red}{RED}{Random Early Detection}
\newacronym{renew}{RENEW}{Reconfigurable Eco-system for Next-generation End-to-end Wireless}
\newacronym{rf}{RF}{Radio Frequency}
\newacronym{rfc}{RFC}{Request for Comments}
\newacronym{rfr}{RFR}{Random Forest Regressor}
\newacronym{ric}{RIC}{RAN Intelligent Controller}
\newacronym{rlc}{RLC}{Radio Link Control}
\newacronym{rlf}{RLF}{Radio Link Failure}
\newacronym{rlnc}{RLNC}{Random Linear Network Coding}
\newacronym{rmr}{RMR}{RIC Message Router}
\newacronym{rmse}{RMSE}{Root Mean Squared Error}
\newacronym{rnis}{RNIS}{Radio Network Information Service}
\newacronym{rr}{RR}{Round Robin}
\newacronym{rrc}{RRC}{Radio Resource Control}
\newacronym{rrm}{RRM}{Radio Resource Management}
\newacronym{rru}{RRU}{Remote Radio Unit}
\newacronym{rs}{RS}{Remote Server}
\newacronym{rsrp}{RSRP}{Reference Signal Received Power}
\newacronym{rsrq}{RSRQ}{Reference Signal Received Quality}
\newacronym{rss}{RSS}{Received Signal Strength}
\newacronym{rssi}{RSSI}{Received Signal Strength Indicator}
\newacronym{rt}{RT}{Real-time}
\newacronym{rtt}{RTT}{Round Trip Time}
\newacronym{ru}{RU}{Radio Unit}
\newacronym{rw}{RW}{Receive Window}
\newacronym{rx}{RX}{Receiver}
\newacronym{s1ap}{S1AP}{S1 Application Protocol}
\newacronym{sa}{SA}{standalone}
\newacronym{sack}{SACK}{Selective Acknowledgment}
\newacronym{sap}{SAP}{Service Access Point}
\newacronym{sc2}{SC2}{Spectrum Collaboration Challenge}
\newacronym{scef}{SCEF}{Service Capability Exposure Function}
\newacronym{sch}{SCH}{Secondary Cell Handover}
\newacronym{scoot}{SCOOT}{Split Cycle Offset Optimization Technique}
\newacronym{sctp}{SCTP}{Stream Control Transmission Protocol}
\newacronym{sdap}{SDAP}{Service Data Adaptation Protocol}
\newacronym{sdk}{SDK}{Software Development Kit}
\newacronym{sdm}{SDM}{Space Division Multiplexing}
\newacronym{sdma}{SDMA}{Spatial Division Multiple Access}
\newacronym{sdn}{SDN}{Software-defined Networking}
\newacronym{sdr}{SDR}{Software-defined Radio}
\newacronym{seba}{SEBA}{SDN-Enabled Broadband Access}
\newacronym{sgsn}{SGSN}{Serving GPRS Support Node}
\newacronym{sgw}{SGW}{Service Gateway}
\newacronym{si}{SI}{Study Item}
\newacronym{sib}{SIB}{Secondary Information Block}
\newacronym{sinr}{SINR}{Signal to Interference plus Noise Ratio}
\newacronym{sip}{SIP}{Session Initiation Protocol}
\newacronym{siso}{SISO}{Single Input, Single Output}
\newacronym{sla}{SLA}{Service Level Agreement}
\newacronym{sm}{SM}{Service Model}
\newacronym{smf}{SMF}{Session Management Function}
\newacronym{smo}{SMO}{Service Management and Orchestration}
\newacronym{sms}{SMS}{Short Message Service}
\newacronym{smsgmsc}{SMS-GMSC}{\gls{sms}-Gateway}
\newacronym{snr}{SNR}{Signal-to-Noise-Ratio}
\newacronym{son}{SON}{Self-Organizing Network}
\newacronym{sptcp}{SPTCP}{Single Path TCP}
\newacronym{srb}{SRB}{Service Radio Bearer}
\newacronym{srn}{SRN}{Standard Radio Node}
\newacronym{srs}{SRS}{Sounding Reference Signal}
\newacronym{ss}{SS}{Synchronization Signal}
\newacronym{sss}{SSS}{Secondary Synchronization Signal}
\newacronym{st}{ST}{Spanning Tree}
\newacronym{svc}{SVC}{Scalable Video Coding}
\newacronym{tb}{TB}{Transport Block}
\newacronym{tcp}{TCP}{Transmission Control Protocol}
\newacronym{tdd}{TDD}{Time Division Duplexing}
\newacronym{tdl}{TDL}{Tapped Delay Line}
\newacronym{tdm}{TDM}{Time Division Multiplexing}
\newacronym{tdma}{TDMA}{Time Division Multiple Access}
\newacronym{tfl}{TfL}{Transport for London}
\newacronym{tfrc}{TFRC}{TCP-Friendly Rate Control}
\newacronym{tft}{TFT}{Traffic Flow Template}
\newacronym{tgen}{TGEN}{Traffic Generator}
\newacronym{tip}{TIP}{Telecom Infra Project}
\newacronym{tm}{TM}{Transparent Mode}
\newacronym{to}{TO}{Telco Operator}
\newacronym{tr}{TR}{Technical Report}
\newacronym{trp}{TRP}{Transmitter Receiver Pair}
\newacronym{ts}{TS}{Technical Specification}
\newacronym{tti}{TTI}{Transmission Time Interval}
\newacronym{ttt}{TTT}{Time-to-Trigger}
\newacronym{tx}{TX}{Transmitter}
\newacronym{uas}{UAS}{Unmanned Aerial System}
\newacronym{uav}{UAV}{Unmanned Aerial Vehicle}
\newacronym{udm}{UDM}{Unified Data Management}
\newacronym{udp}{UDP}{User Datagram Protocol}
\newacronym{udr}{UDR}{Unified Data Repository}
\newacronym{ue}{UE}{User Equipment}
\newacronym{uhd}{UHD}{\gls{usrp} Hardware Driver}
\newacronym{ul}{UL}{Uplink}
\newacronym{um}{UM}{Unacknowledged Mode}
\newacronym{umi}{UMi}{Urban Micro}
\newacronym{uml}{UML}{Unified Modeling Language}
\newacronym{upa}{UPA}{Uniform Planar Array}
\newacronym{upf}{UPF}{User Plane Function}
\newacronym{urllc}{URLLC}{Ultra Reliable and Low Latency Communications}
\newacronym{usa}{U.S.}{United States}
\newacronym{usim}{USIM}{Universal Subscriber Identity Module}
\newacronym{usrp}{USRP}{Universal Software Radio Peripheral}
\newacronym{utc}{UTC}{Urban Traffic Control}
\newacronym{vim}{VIM}{Virtualization Infrastructure Manager}
\newacronym{vm}{VM}{Virtual Machine}
\newacronym{vnf}{VNF}{Virtual Network Function}
\newacronym{volte}{VoLTE}{Voice over \gls{lte}}
\newacronym{voltha}{VOLTHA}{Virtual OLT HArdware Abstraction}
\newacronym{vr}{VR}{Virtual Reality}
\newacronym{vran}{vRAN}{Virtualized \gls{ran}}
\newacronym{vss}{VSS}{Video Streaming Server}
\newacronym{wbf}{WBF}{Wired Bias Function}
\newacronym{wf}{WF}{Waterfilling}
\newacronym{wg}{WG}{Working Group}
\newacronym{wi}{WI}{Wireless InSite}
\newacronym{wlan}{WLAN}{Wireless Local Area Network}
\newacronym{osm}{OSM}{Open Source \gls{nfv} Management and Orchestration}
\newacronym{pnf}{PNF}{Physical Network Function}
\newacronym{mtc}{MTC}{Machine-type Communications}
\newacronym{mns}{MnS}{Management Services}
\newacronym{ves}{VES}{\gls{vnf} Event Stream}
\newacronym{ei}{EI}{Enrichment Information}
\newacronym{fh}{FH}{Fronthaul}
\newacronym{fft}{FFT}{Fast Fourier Transform}
\newacronym{laa}{LAA}{Licensed-Assisted Access}
\newacronym{plfs}{PLFS}{Physical Layer Frequency Signals}
\newacronym{ptp}{PTP}{Precision Time Protocol}
\newacronym{cbrs}{CBRS}{Citizen Broadband Radio Service}
\newacronym{otic}{OTIC}{Open Testing and Integration Center}
\newacronym{sba}{SBA}{Service-Based Architecture}
\newacronym{cif}{CI}{cyberinfrastructure}
\newacronym{sonic}{SONiC}{Software for Open Networking in the Cloud}
\newacronym{ocp}{OCP}{Open Compute Project}
\newacronym{snmp}{SNMP}{Simple Network Management Protocol}
\newacronym{raid}{RAID}{redundant array of independent disks}
\newacronym{nfs}{NFS}{Network File Storage}
\newacronym{ci}{CI}{Continuous Integration}
\newacronym{cd}{CD}{Continuous Deployment}
\newacronym{dtn}{DTN}{Data Transfer Node}

\newacronym{dns}{DNS}{Domain Name Service}
\newacronym{nrpe}{NRPE}{Nagios Remote Plugin Executor}
\newacronym{ldap}{LDAP}{Lightweight Directory Access Protocol}
\newacronym{lan}{LAN}{Local Area Network}
\newacronym{vlan}{VLAN}{Virtual LAN}

\newacronym{ipmi}{IPMI}{Intelligent Platform Management Interface}
\newacronym{tor}{ToR}{Top-of-the-Rack}
\newacronym{lmn}{LMN}{Large Memory Node}
\newacronym{bgp}{BGP}{Border Gateway Protocol}
\newacronym{dhcp}{DHCP}{Dynamic Host Configuration Protocol}
\newacronym{vrf}{VRF}{Virtual Routing and Forwarding}
\newacronym{vpn}{VPN}{Virtual Private Network}
\newacronym{rma}{RMA}{Return Merchandise Authorization}
\newacronym{hpc}{HPC}{High Performance Compute}

\newacronym{nu}{NU}{Northeastern University}
\newacronym{asic}{ASIC}{Application-specific Integrated Circuit}
\newacronym{rdma}{RDMA}{Remote Direct Memory Access}
\newacronym{roce}{RoCE}{RDMA over Converged Ethernet}
\newacronym{ovs}{OVS}{Open vSwitch}
\newacronym{frr}{FRR}{Free Range Routing}
\newacronym{ups}{UPS}{Uninterruptible Power Supply}

\newacronym{ntia}{NTIA}{National Telecommunications and Information Administration}
\newacronym{pii}{PII}{Personal and Identifiable Information}
\newacronym{irb}{IRB}{Institutional Review Board}
\newacronym{doi}{DOI}{Digital Object Identifier}

\newacronym{sdo}{SDO}{Standard-Development Organization}
\newacronym{ose}{OSE}{Open Source Ecosystem}
\newacronym{osc}{OSC}{O-RAN Software Community}
\newacronym{dop}{DOP}{Director of Operations}
\newacronym{pm}{PM}{Program Manager}
\newacronym{excom}{EXCOM}{Executive Committee}
\newacronym{iiot}{IIoT}{Industrial \gls{iot}}
\newacronym{lf}{LF}{Linux Foundation}

\newacronym{wiot}{WIoT}{Institute for the Wireless Internet of Things}
\newacronym{rl}{RL}{Reinforcement Learning}
\newacronym{drl}{DRL}{Deep Reinforcement Learning}

\newacronym{nofo}{NOFO}{Notice of Funding Opportunity}

\newacronym{onr}{ONR}{Office of Naval Research}
\newacronym{afosr}{AFOSR}{Air Force Office of Scientific Research}
\newacronym{afrl}{AFRL}{Air Force Research Laboratory}
\newacronym{arl}{ARL}{Army Research Laboratory}

\newacronym{arc}{ARC}{Aerial Research Cloud}
\newacronym{cast}{CaST}{Channel emulation scenario generator and Sounder Toolchain}

\newacronym{mno}{MNO}{Mobile Network Operator}
\newacronym{ct}{CT}{Continuous Testing}
\newacronym{oci}{OCI}{Open Container Initiative}

\newacronym{xai}{XAI}{Explainable AI}
\newacronym{sas}{SAS}{Spectrum Access System}

\newacronym{rem}{REM}{Random Ensemble Mixture}
\newacronym{ns3}{ns-3}{Network Simulator 3}

\newacronym{fcu}{FCU}{Flight Control Unit}
\newacronym{ros}{ROS}{Robot Operating System}
\newacronym{c2}{C2}{Command and Control}
\newacronym{esc}{ESC}{Electronic Speed Controller}
\newacronym{blos}{BLoS}{Beyond-Line-of-Sight}
\tikzstyle{startstop} = [rectangle, rounded corners, minimum width=2cm, minimum height=0.5cm,text centered, draw=black]
\tikzstyle{io} = [trapezium, trapezium left angle=70, trapezium right angle=110, minimum width=3cm, minimum height=1cm, text centered, draw=black]
\tikzstyle{process} = [rectangle, minimum width=2cm, minimum height=0.5cm, text centered, draw=black, alignb=center]
\tikzstyle{decision} = [ellipse, minimum width=2cm, minimum height=1cm, text centered, draw=black]
\tikzstyle{arrow} = [thick,<->,>=stealth]
\tikzstyle{line} = [thick,>=stealth]
\tikzstyle{darrow} = [thick,<->,>=stealth,dashed]
\tikzstyle{sarrow} = [thick,->,>=stealth]
\tikzstyle{larrow} = [line width=0.3mm,dashdotted,->,>=stealth]
\tikzstyle{llarrow} = [line width=0.1mm,->,>=stealth]

\makeatletter
\def\grd@save@target#1{%
  \def\grd@target{#1}}
\def\grd@save@start#1{%
  \def\grd@start{#1}}
\tikzset{
  grid with coordinates/.style={
    to path={%
      \pgfextra{%
        \edef\grd@@target{(\tikztotarget)}%
        \tikz@scan@one@point\grd@save@target\grd@@target\relax
        \edef\grd@@start{(\tikztostart)}%
        \tikz@scan@one@point\grd@save@start\grd@@start\relax
        \draw[minor help lines] (\tikztostart) grid (\tikztotarget);
        \draw[major help lines] (\tikztostart) grid (\tikztotarget);
        \grd@start
        \pgfmathsetmacro{\grd@xa}{\the\pgf@x/1cm}
        \pgfmathsetmacro{\grd@ya}{\the\pgf@y/1cm}
        \grd@target
        \pgfmathsetmacro{\grd@xb}{\the\pgf@x/1cm}
        \pgfmathsetmacro{\grd@yb}{\the\pgf@y/1cm}
        \pgfmathsetmacro{\grd@xc}{\grd@xa + \pgfkeysvalueof{/tikz/grid with coordinates/major step x}}
        \pgfmathsetmacro{\grd@yc}{\grd@ya + \pgfkeysvalueof{/tikz/grid with coordinates/major step y}}
        \foreach \x in {\grd@xa,\grd@xc,...,\grd@xb}
        \node[anchor=north] at (\x,\grd@ya) {\pgfmathprintnumber{\x}};
        \foreach \y in {\grd@ya,\grd@yc,...,\grd@yb}
        \node[anchor=east] at (\grd@xa,\y) {\pgfmathprintnumber{\y}};
      }
    }
  },
  minor help lines/.style={
    help lines,
    gray,
    line cap =round,
    xstep=\pgfkeysvalueof{/tikz/grid with coordinates/minor step x},
    ystep=\pgfkeysvalueof{/tikz/grid with coordinates/minor step y}
  },
  major help lines/.style={
    help lines,
    line cap =round,
    line width=\pgfkeysvalueof{/tikz/grid with coordinates/major line width},
    xstep=\pgfkeysvalueof{/tikz/grid with coordinates/major step x},
    ystep=\pgfkeysvalueof{/tikz/grid with coordinates/major step y}
  },
  grid with coordinates/.cd,
  minor step x/.initial=.5,
  minor step y/.initial=.2,
  major step x/.initial=1,
  major step y/.initial=1,
  major line width/.initial=1pt,
}
\makeatother

\usepackage{dblfloatfix}

\begin{document}

\title{5G Aero: A Prototyping Platform for Evaluating Aerial 5G Communications}
%OR
% Optimizing Compact UAV Performance with 5G Integration for Advanced Telecommunications

% Mavenir Team is Sai Satish, Manoj AnanthaSwamy Nittoor, Rajarajan Sivaraj.

\author{\IEEEauthorblockN{Matteo Bordin, Madhukara S. Holla, Sakthivel Velumani, Salvatore D'Oro, Tommaso Melodia}
\IEEEauthorblockN{Institute for the Wireless Internet of Things, Northeastern University, Boston, MA, USA\\
Email: \{{bordin.m, sholla.m, s.velumani, s.doro, t.melodia\}@northeastern.edu}\\
}
\thanks{This work was partially supported by the National Telecommunications and Information Administration (NTIA)'s Public Wireless Supply Chain Innovation Fund (PWSCIF) under Award No. 25-60-IF002.}
}

\maketitle

\begin{abstract}
The application of small-factor, 5G-enabled \glspl{uav} has recently gained significant interest in various aerial and Industry 4.0 applications. However, ensuring reliable, high-throughput, and low-latency 5G communication in aerial applications remains a critical and underexplored problem. This paper presents the \gls{5g} Aero, a compact \gls{uav} optimized for \gls{5g} connectivity, aimed at fulfilling stringent \gls{3gpp} requirements. We conduct a set of experiments in an indoor environment, evaluating the \gls{uav}'s ability to establish high-throughput, low-latency communications in both \gls{los} and \gls{nlos} conditions. Our findings demonstrate that the \gls{5g} Aero meets the required \gls{3gpp} standards for \gls{c2} packets latency in both \gls{los} and \gls{nlos}, and video latency in \gls{los} communications and it maintains acceptable latency levels for video transmission in \gls{nlos} conditions. Additionally, we show that the 5G module installed on the \gls{uav} introduces a negligible 1\% decrease in flight time, showing that \gls{5g} technologies can be integrated into commercial off-the-shelf  \glspl{uav} with minimal impact on battery lifetime. This paper contributes to the literature by demonstrating the practical capabilities of current \gls{5g} networks to support advanced \gls{uav} operations in telecommunications, offering insights into potential enhancements and optimizations for \gls{uav} performance in \gls{5g} networks. 
\end{abstract}

\glsresetall
\glsunset{ns3}
\glsunset{nr}
\glsunset{lte}

\vspace{-0.2cm}
\section{Introduction}
\label{sec:intro}
Over the past decade, \glspl{uav} have evolved from niche technologies into indispensable tools across various industries. They are utilized in agriculture, logistics, and other sectors to collect data, transport goods, and handle tasks that are hazardous or time-consuming for humans \cite{agricolture,dataFromSensor}. However, one of the most promising applications of \glspl{uav} is tied to the telecommunications field, where drones can serve as mobile base stations, signal relays, or aerial data collection platforms \cite{aerialBs,drivongFromThesky}. 

With the advent of \gls{5g} technology, the role of \glspl{uav} in telecommunications is more significant than ever.
\gls{5g} networks support \gls{urllc} and \gls{embb}, enabling \gls{uav} operations that demand ultra-reliable communications and high data rates for real-time monitoring, asset tracking, disaster recovery, and autonomy. Additionally, \gls{5g} facilitates multi-user communication over large areas, unlocking new opportunities for drone swarms and \gls{uav} applications in complex indoor Industry 4.0 environments \cite{lowLatency}.

\textbf{Motivation:} Despite significant advancements in \gls{uav} technologies, there remains a substantial gap in the development of compact and agile \gls{uav} platforms with \gls{5g} capabilities for telecommunication research. Not only does the consumer drone market offer relatively few options, but it is also often expensive and lacks the necessary open \glspl{api} and configurable \gls{5g} modules. Furthermore, existing \glspl{uav} designed for \gls{5g} research tend to be large or prohibitively costly, making them unsuitable for constrained environments such as indoor spaces and narrow outdoor areas. This limitation highlights the need for an open-source, affordable, and customizable \gls{uav} platform tailored for high-throughput, low-latency \gls{5g} research.

Moreover, to the best of our knowledge, no comprehensive analyses exist on how \gls{5g} connectivity impacts \gls{uav} performance, particularly in terms of packet transmission latency and energy consumption. The \gls{3gpp} standard defines strict latency requirements for both \gls{los} and \gls{nlos} applications. For instance, in \gls{los} scenarios, uplink \gls{c2} packet latency must be under 40ms, while video packet latency must remain below 1s. In \gls{nlos} conditions, the latency requirement for video packets is even more stringent, at under 140ms \cite{ETSI181} \cite{ETSI182}.  Meeting these requirements is not just crucial for current \gls{uav} operations but also paves the way for \gls{blos} flights, where UAVs must operate without direct visual contact with the pilot. Without ensuring reliable \gls{5g} connectivity that meets \gls{los} and \gls{nlos} latency constraints, achieving safe and effective \gls{blos} operations will remain unattainable.

\textbf{Contributions:} In this paper, we make several key contributions.  Firstly, we present a detailed workflow for updating an outdated commercial drone (Intel Aero) into a fully functional \gls{5g}-enabled \gls{uav}, creating a cost-effective, small-form-factor research platform for studying aerial \gls{5g} networks.
Secondly, we introduce the \gls{5g} Aero, a prototyping platform designed to evaluate whether existing \gls{5g} technologies can support UAV communications while meeting \gls{3gpp} latency and throughput benchmarks in both \gls{los} and \gls{nlos} conditions.
Finally, we assess the \gls{5g} Aero’s performance under diverse scenarios and configurations to determine whether current \gls{5g} and \gls{uav} technologies available to the research community (e.g., \gls{5g} \gls{ran} and core network software, flight controllers and \gls{5g} modems) comply with \gls{3gpp}-specified benchmarks for latency and throughput in real-world \gls{los} and \gls{nlos} conditions. Additionally, we evaluate the impact of \gls{5g} connectivity on the \gls{uav} flight duration. 

% The development of such platform not only tests the capabilities of existing 5G technologies in novel applications but also provides a foundation for future advancements in UAV and telecommunications integration.
The remainder of the paper is organized as follows: Section \ref{sec:soa} reviews the current state of the art in \gls{uav} for \gls{5g} applications, highlighting the limitations and potential of existing platforms. Section \ref{sec:design} describes the methodology and technical modifications required to upgrade a discontinued drone into the \gls{5g} Aero, including hardware enhancements and \gls{5g} integration. Section \ref{sec:experiments} details the experimental setup, outlining the testing environment, evaluation metrics, and procedures for assessing the \gls{uav}'s performance under \gls{los} and \gls{nlos} conditions. Section \ref{sec:results} presents the results, analyzing the \gls{uav}’s battery life, latency and reliability against \gls{3gpp} benchmarks. Section \ref{sec:challenges} discusses the challenges encountered during development and testing. Finally, section \ref{sec:conclusions} concludes the paper with a summary of our key findings.

%such as improving the drone with self-navigating abilities by incorporating new sensors.

\section{State of the Art}
\label{sec:soa}
% talk about Drone competitors
The integration of \glspl{uav} with \gls{5g} technologies is an emerging field that demands careful consideration of \gls{uav} capabilities, size, payload, and cost. 
% Several \gls{uav} platforms have been utilized in research to explore their feasibility in \gls{5g} communication environments, each with distinct features and trade-offs.
Monarch is a quadcopter equipped with a companion computer and a \gls{usrp} that offers a programmable \gls{uav} platform to operate in the sub-6GHz frequency range~\cite{Buczek_2021}. Thanks to the use of \gls{sdr}, Monarch enables complex wireless research with diverse wireless technologies and standards. However, its large size limits its application to outdoor scenarios and prevents its use for indoor applications.
In contrast, the AIRLink \cite{airlink} is a compact \gls{fcu} featuring a \gls{5g} connectivity module to support low latency and high data bandwidth. It supports up to 600 Mbps bandwidth and operates on \gls{5g} sub-6 and mmWave frequencies. However, despite the many features available in this small form-factor product, the AIRLink module only offers the \gls{fcu} and networking capabilities, which needs to be integrated with a \gls{uav}, making this solution relatively expensive.
In \cite{ferranti2020skycell}, authors used the DJI Matrice 600, a hexacopter equipped with a \gls{usrp}, to act as an aerial \gls{gnb}. The \gls{uav} offers significant payload capacity, making it suitable for carrying heavy equipment. However, its high cost and large size make the platform expensive and unsuitable for flying in indoor environments.
A small and affordable \gls{uav} alternative is the Crazyflie 2.1. The Crazyflie has been considered in \cite{nanaoUAV} and is a quadrotor of about 12 cm by 12 cm (including the propellers). It has been used to fly in small tunnels but it is not capable of holding a \gls{5g} radio. This limitation restricts its potential for \gls{5g}-based communication research, highlighting the trade-off between size, cost, and technological capability.
Finally, the Sky5G \gls{uav}~\cite{1mDrone} integrates \gls{5g} capabilities through a \gls{usrp} mini and a form factor characterized by a 1-meter diagonal. Sky5G acts as an aerial base station, but does not offer \gls{ue} capabilities. 
% In summary, the aforementioned \gls{uav} platforms demonstrate a wide range of trade-offs in terms of size, cost, and functionality. While larger drones like the DJI Matrice 600 and Monarch are capable of handling heavy payloads and supporting advanced communication technologies, they come with high costs and size constraints. More compact drones, such as the AIRlink module and the Sky5G, offer \gls{5g} integration but at a higher price, while smaller, cheaper drones sacrifice \gls{5g} capabilities for affordability. 
In this paper, we fill the gap in the literature by presenting 5G Aero, a compact, cost-effective \gls{5g}-enabled custom \gls{ue} drone designed specifically for indoor applications. Unlike typical \gls{5g} drones that are often bulky and expensive, our design ensures a compact size for flying in confined spaces without compromising performance. Moreover, unlike previous works, we evaluate the \gls{5g} \gls{uav} performance by assessing whether it can meet the stringent requirements defined by \gls{3gpp}.
% By achieving this balance, we set a new benchmark for compact \gls{5g} custom drones, paving the way for more accessible and versatile \gls{uav} solutions in various sectors.

\section{5G Aero Design and Prototype}
\label{sec:design} 
To build a compact and \gls{5g} capable drone, it is important to understand its architecture and the core components that dictate its functionality and performance. A \gls{uav} consists of several key elements. The frame provides structural support and houses all components securely. The computing unit handles data processing, communication, and control tasks. The \gls{fcu} stabilizes the drone, processes sensor inputs, and executes flight commands. The radio front-ends enable wireless communication. Additionally, the \gls{uav} is battery powered, it has propellers driven by \glspl{esc} and it can incorporate sensors (e.g. LiDAR, and optical flow) for navigation and environmental awareness.

%Specifically, the main components of a \gls{uav} are the \textit{frame}, the \textit{computing unit}, the \textit{\gls{fcu}}, and the \textit{radio front-ends}. 
% and adheres to specific design principles, such as implementing the \textit{wireless stack} in the computing unit and exposing \textit{control APIs} through Mavlink or \gls{ros}. 

% Three main principles guided the development of this drone: 
% i) Designing a \gls{uav} with \gls{5g} connectivity and strong computing capabilities, with the core components mentioned before.
% ii) Ensuring the drone is open-source and easy to upgrade.
% iii) Making the \gls{uav} compact for indoor flight.
% iv) Keeping the project low-cost and simple so others can replicate it.
% \vspace{-0.055cm}
In the following, we first describe the baseline \gls{uav} platform we used in this work (i.e., the Intel Aero), and then provide a detailed description of how we updated the original design to integrate \gls{5g} capabilities. Specifically, our design has been engineered to meet the following core principles: (i) support for \gls{5g} \gls{ue} operations so that the \gls{uav} can act as an aerial mobile terminal; (ii) programmability and support for future extensions; (iii) small form-factor; (iv) contained cost.

\begin{figure*}[t!]
    \centering
    \begin{subfigure}[t]{\columnwidth}
        \centering
    \vspace{-4.77cm}
        \includegraphics[width=0.8\textwidth]{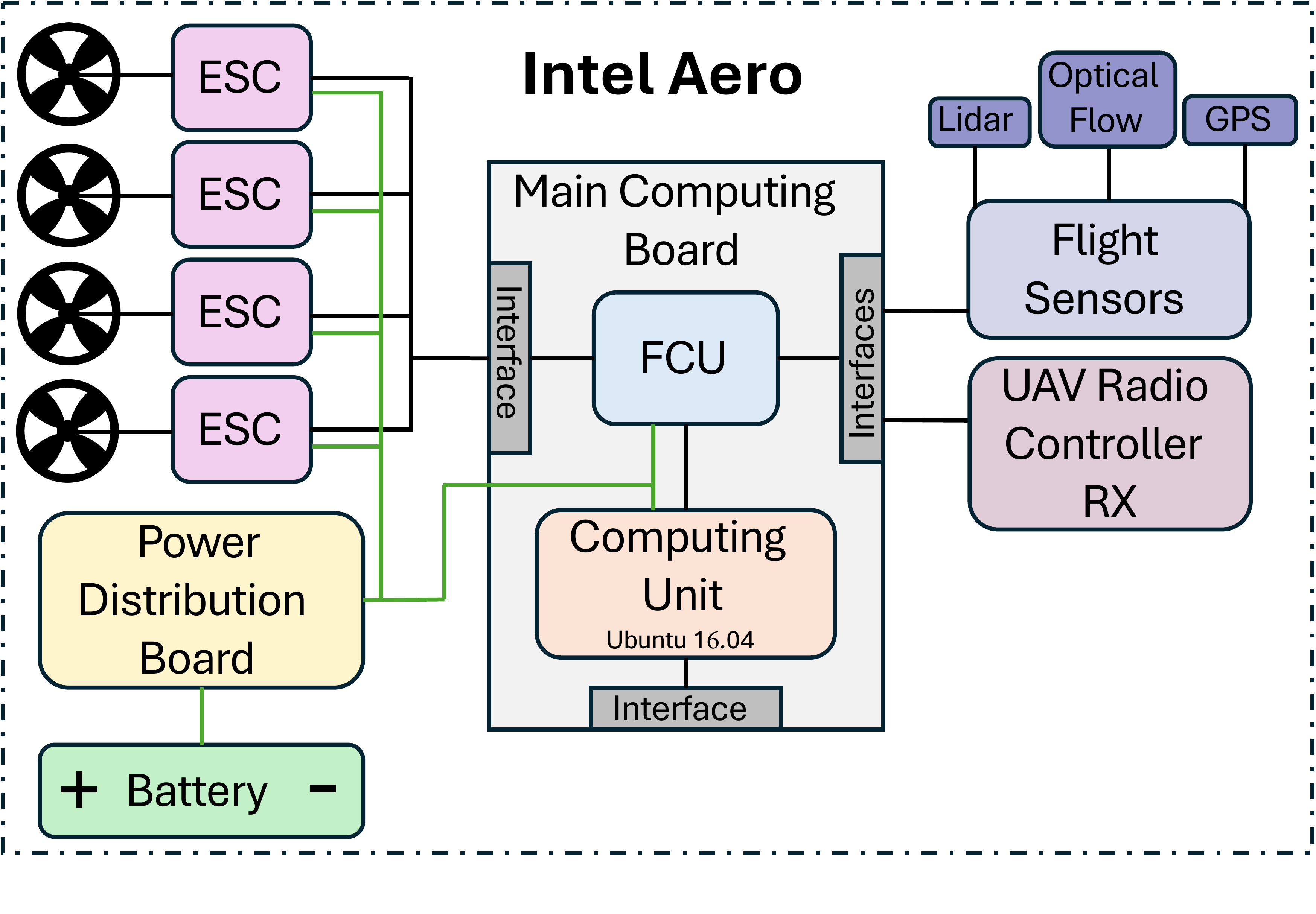}
        \vspace{-0.2cm}
        \caption{Intel Aero design.}
        \label{fig:Aero}
            \vspace{-0.1cm}
    \end{subfigure}
    \hfill % Adds horizontal space between the figures
    \begin{subfigure}[t]{\columnwidth}
        \centering
        \includegraphics[width=0.75\textwidth]{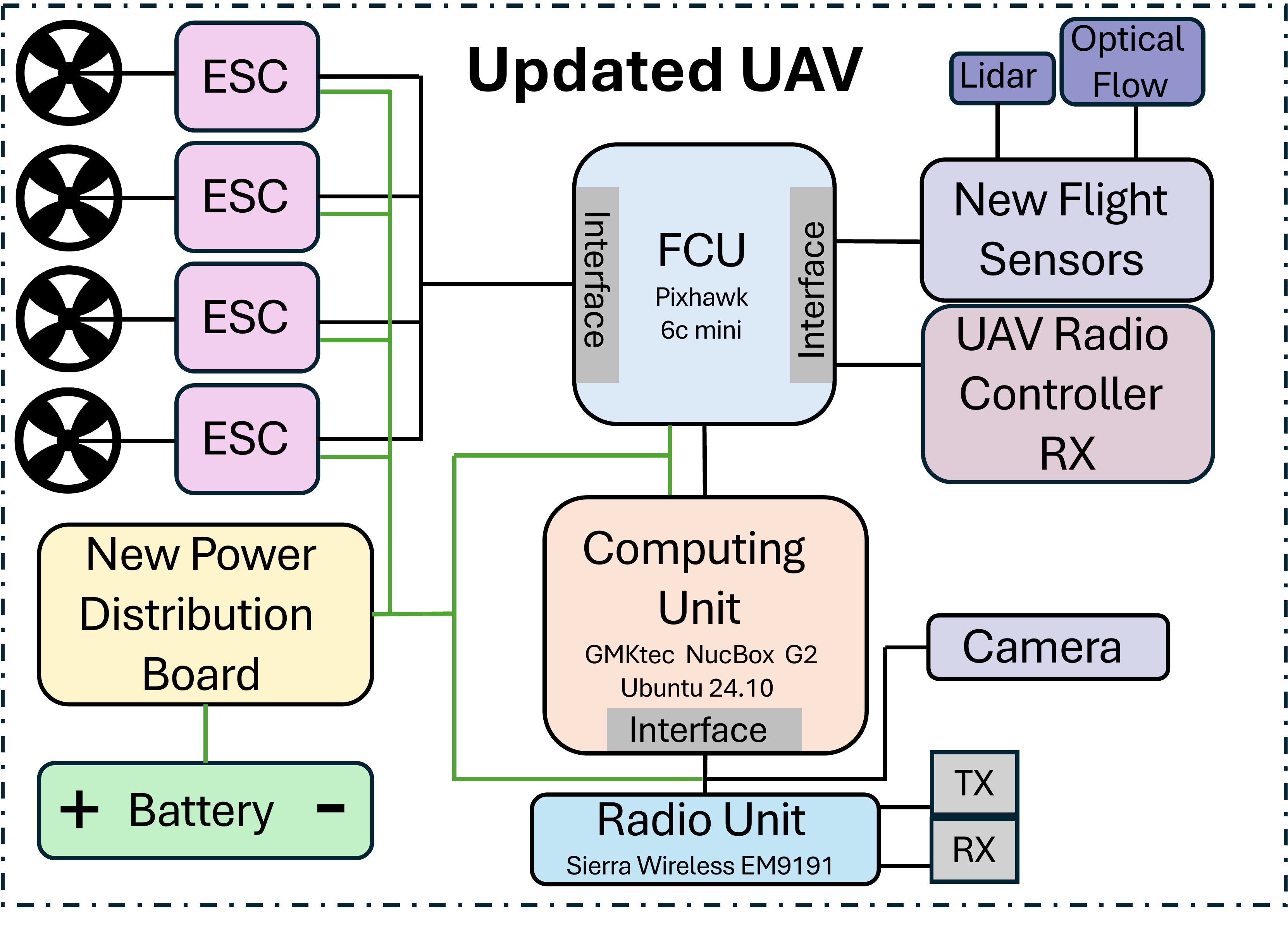}
        \caption{Updated 5G Aero design.}
        \label{fig:Updated}
            \vspace{-0.1cm}
    \end{subfigure}
    \caption{Architectural view of the Intel Aero and the 5G Aero.}
    \label{fig:SystArchit}
    \vspace{-0.3cm}
\end{figure*}

\textbf{Baseline Platform}
With these design considerations in mind, we selected the Intel Aero as the baseline platform. 
% looked up for an existing \gls{uav} platform that could serve as a foundation for our custom \gls{5g}-enabled drone while minimizing development overhead. 
The Intel Aero is a discontinued UAV platform designed to provide a comprehensive development kit for \gls{uav} technology enthusiasts and professionals. Central to its architecture, there is a computing unit running Ubuntu 16, a dedicated \gls{fcu}, and a versatile frame equipped with essential \gls{uav} components such as a GPS. The platform became popular for supporting a range of sensors and accessories aimed to enhance \gls{uav} capabilities. 
% However, in 2019, production of the Intel Aero was discontinued, leading to the cessation of official support from the manufacturer. 
% As of 2025, attempts to update the kernel to install more recent software versions have resulted in critical failures, including the inability of the drone to power on.
Despite its obsolete operating system, outdated drivers, and lack of technical support and \gls{5g} compatibility, many of its components—such as the frame and motors—still meet the required size and suitability for indoor applications. For this reason, we decided to remove the obsolete elements from the Intel Aero and re-purpose its frame and rotors.
% Intel Aero 2 Deisgn
Fig.~\ref{fig:Aero} shows how the original Intel Aero was designed to minimize component count. For this reason, it featured an integrated main computing board hosting both the \gls{fcu} and the Computing Unit. This integration indeed helps reduce the form factor but limits the ability to upgrade the design and add or remove components. The main computing board is powered by a power distribution board which, draws energy from a battery placed in a dedicated slot of the frame. The power distribution board also supplies power to the motors through \glspl{esc}, which connect to the \gls{fcu} via a dedicated interface on the main computing board. Additional interfaces on the \gls{fcu} allows the connections of external sensors such as GPS, lidar, and optical flow, as well as the \gls{uav}’s radio controller receiver. The only interface offered by the Computing Unit is a micro USB B port, necessitating a converter to ensure compatibility with other connectors. This port must be split with additional extenders to connect more than one device to the Computing Unit. Finally, the \gls{fcu} communicates with all the components through MAVLink messaging protocol, while the Computing Unit, which runs an outdated Linux version, directly interacts with the \gls{fcu} through MAVLink Router, facilitating direct interactions with \gls{uav} telemetry.
%TALK ONLY ABOUT THE GENERAL STRUCTURE OF THE DRONE, SO THE DESIGN. DO NOT FOCUS ON THE MODEL OF THE COMPONENTS. mAYBE SAY JUST WHICH PIECES WE HAD TO CHANGE (E.G FCU, ESC,..)

\textbf{5G Aero Design} \label{UAV design}
Figure \ref{fig:Updated} illustrates the design of our \gls{5g} Aero. The primary modification is the removal of the main computing board, which facilitates the disaggregation of the \gls{fcu} and the Computing Unit. This separation enables the individual updating of components, allowing for the selection of a computing unit with higher computational power, a GPU, or specific ports. Another significant enhancement is the addition of power connectivity, which supports powering an additional sensor or radio directly from the power distribution board. These two redesigns significantly enhance the platform's versatility. First, the computational limitations are now resolved as the computing unit can be personalized to specific use cases, whether requiring a dedicated GPU, additional memory, or I/O ports. Second, the added power connector allows for the integration of more advanced devices, such as a \gls{5g} radio, expanding the \gls{uav}’s operational capabilities.

\begin{figure}[t!]
    \centering
    \includegraphics[width=0.9\columnwidth]{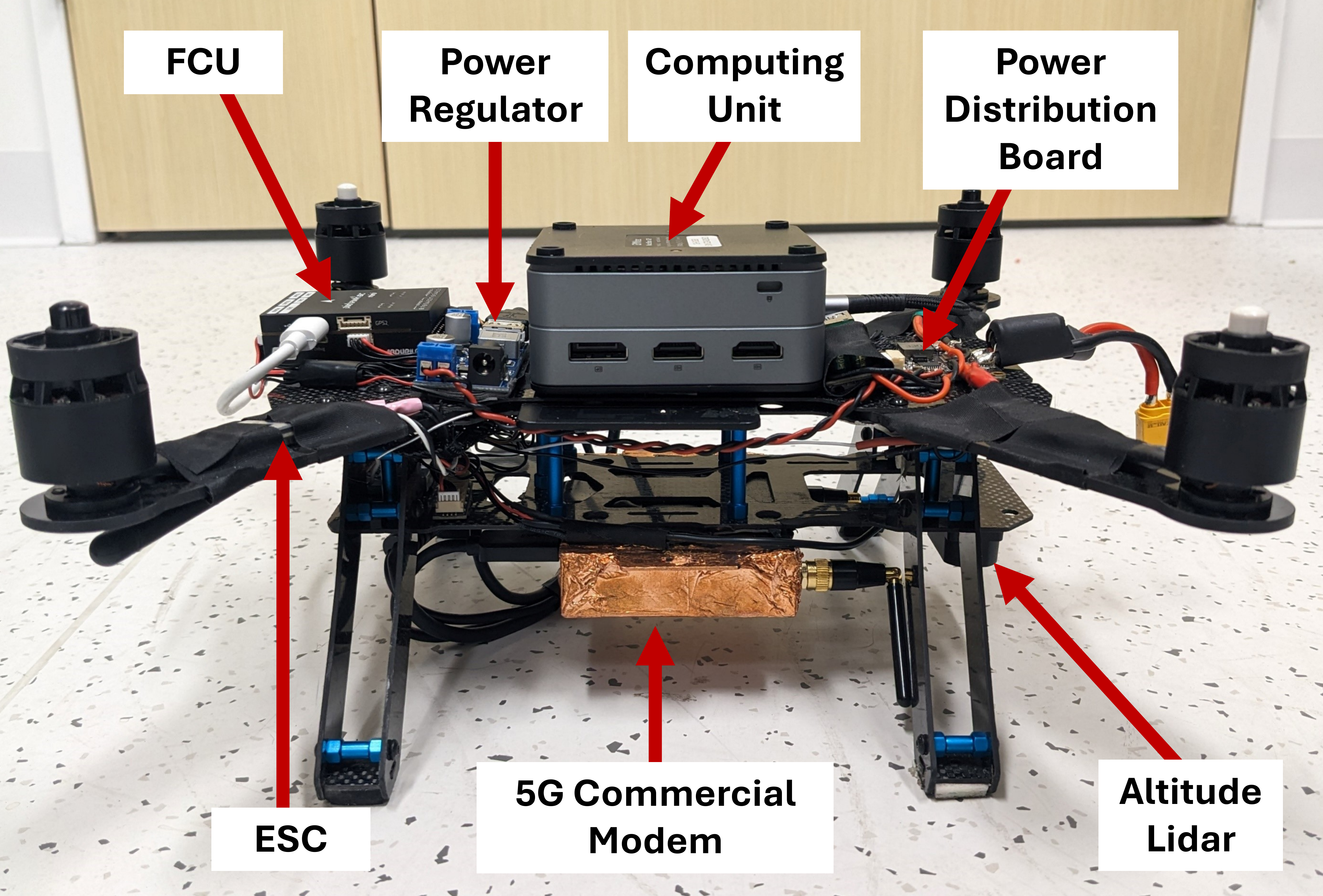}
    \caption{The 5G Aero prototype.}
    \label{fig:MyUAV}
    \vspace{-0.4cm}
\end{figure}

Fig.~\ref{fig:MyUAV} shows the 5G Aero \gls{uav} we have built following the design described above and shown in Fig.~\ref{fig:Updated}. As \gls{fcu} we have installed the latest and smallest model of the Pixhawk family, the 6C mini. We have paired it with the GMKtec NucBox G2, a small factor Computing Unit, 3.3x3.3x1.37 inches (8.5x8.5x3.5 cm), with 12GB of RAM, and an Intel 12th Alder Lake N100 processor, which perfectly fits within the \gls{uav}'s frame. The \gls{uav}’s original \glspl{esc}, which had proprietary protocols incompatible with MAVLink's open protocol DShot, were replaced to enable telemetry data transmission, essential for the \gls{fcu} to optimize stability and altitude control. We have connected a commercial \gls{5g} modem (Sierra Wireless EM9191 NR \gls{5g} Modem) to the Computing Unit and the power distribution board using a suitable power regulator. Additionally, we have connected an altitude lidar to the \gls{fcu} for continuous altitude control during flights. 

\begin{figure}[t!]
    \begin{subfigure}[t]{\columnwidth}
    \centering
        \includegraphics[width=0.9\columnwidth]{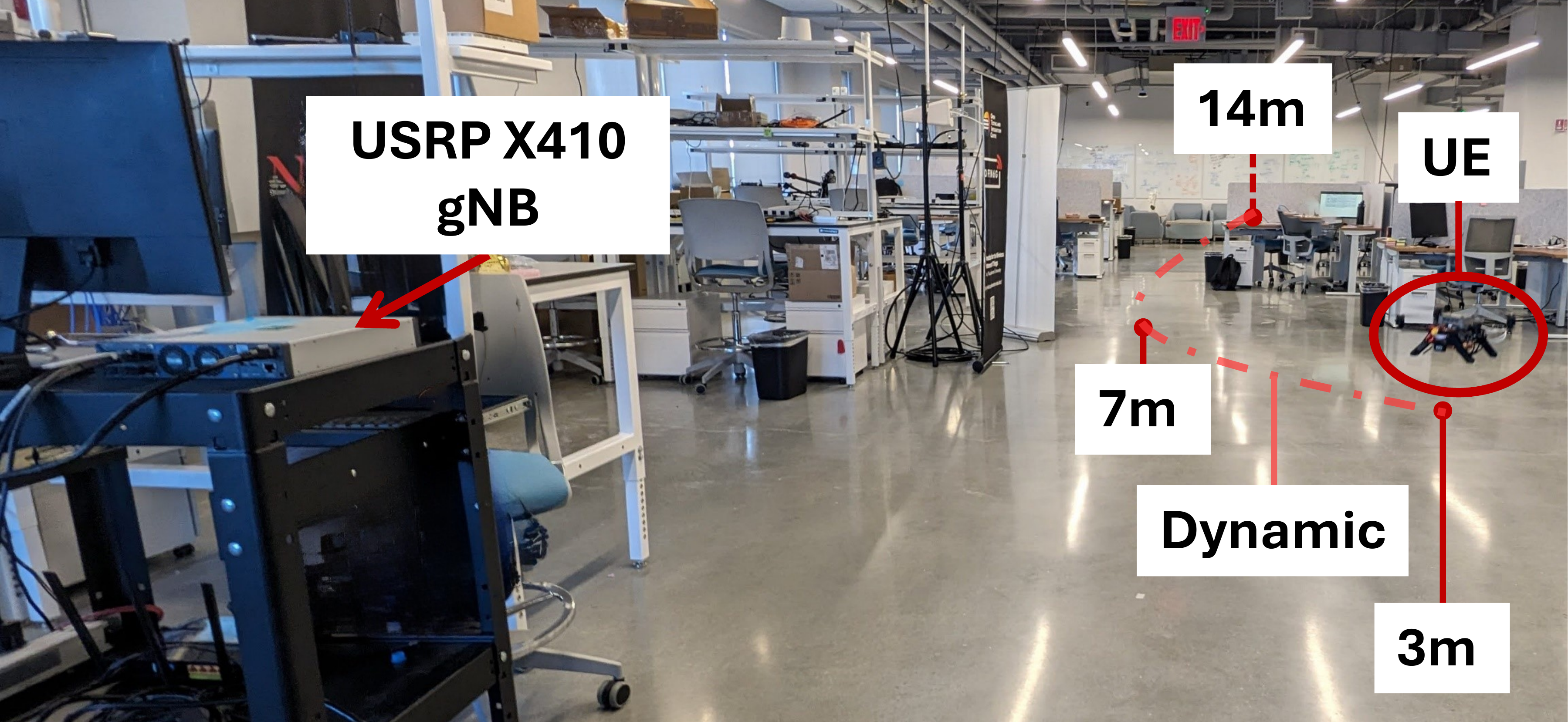}
        \caption{The indoor experimental setup showing the 5G Aero (to the right) and the \gls{gnb} (to the left), and the four scenarios with the corresponding distances.}
        \label{fig:ExpSetup}
    \end{subfigure}
    \begin{subfigure}[t]{\columnwidth}
    \centering
    \vspace{0.3cm}
        \includegraphics[width=0.9\columnwidth]{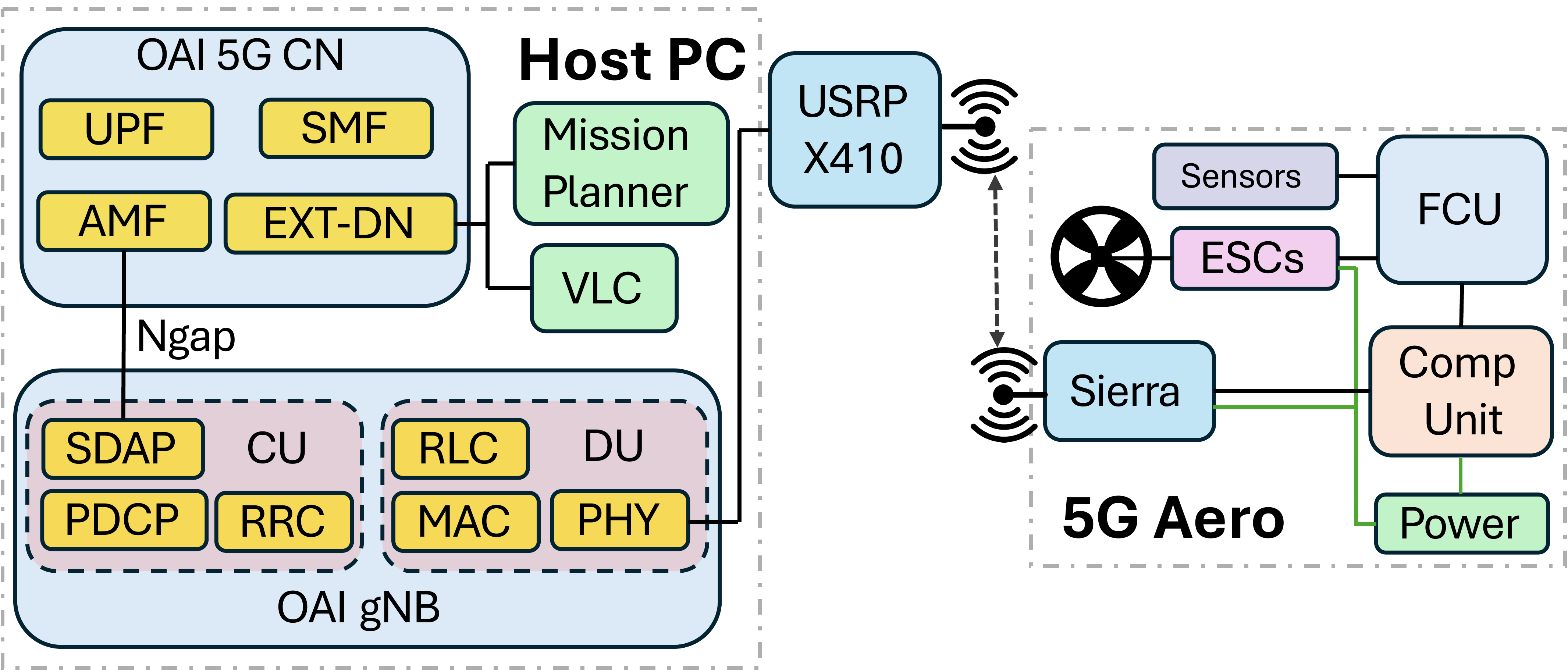}
        \caption{Architectural diagram of the experimental setup.}
        \label{fig:ExpOverview}
    \end{subfigure}
    \caption{Experiment scenario.}
    \label{fig:ExpScen}
    \vspace{-0.5cm}
\end{figure}

MAVLink Router has been installed on the Computing Board to facilitate efficient message management between the \gls{fcu}, drone sensors and ground station. Finally, we implemented the full protocol stack from OpenAirInterface \cite{oai} to establish and maintain a secure \gls{5g} connection between the \gls{uav} and the \gls{gnb} (Fig.~\ref{fig:ExpOverview}). Since we intend to operate this drone exclusively indoors, where GPS is ineffective and adds unnecessary weight, no GPS was installed.

\section{Experiment Setup}
\label{sec:experiments}
%describe the goal of the experiment: use case and what we want to demonstrate
We aim to evaluate 5G Aero's performance in diverse experimental scenarios and evaluate its compliance with the \gls{3gpp}-specified benchmarks for UAV operations and related to latency and reliability. This work specifically focuses on an indoor scenario: an office environment. This setup, described below, contains numerous obstacles, clutter, and reflective objects that can attenuate or disrupt the signal between the \gls{uav} and the \gls{gnb}.
%Should we say again what is the goal?

% describe the drone setup for this experiment. Which sensors were mounted, which model of fcu, computing board, esc,... % describe in details what we removed and why. Say what we bought. 

%\subsection{Experiment Setup}
%creae a better intro talking about the goal of the experiment

\subsection{Experiment Overview} 
To determine if a small form-factor \glspl{uav} equipped with \gls{5g} technology can fulfill the stringent requirements of \gls{uav}-based applications (which will be detailed below), we conducted a flight test in an indoor environment designed to simulate a \gls{uav} flying inside an office or laboratory. This environment enabled \gls{uav} to navigate through \gls{los} and \gls{nlos} spaces, with obstacles made of various materials that influence signal propagation due to multi-path and attenuation. The objective of this experiment is to assess whether the \gls{5g} \gls{uav} could satisfy the \gls{3gpp} flight requirements stipulated in the latest ETSI release \cite{ETSI181,ETSI182}. At least 99.9\% of \gls{c2} packets, both in the uplink and downlink, must have a latency below 40ms. Also, at least 99.9\% of uplink video packets in \gls{los} must have a latency of no more than 1s, while for video in \gls{nlos}, at least 99.99\% of packets must have a latency under 140ms.
% \begin{table}[h]
% \centering
% \begin{tabular}{|c|c|c|c|}
% \hline
% \textbf{Functionlity} & \textbf{Latency} & \textbf{Reliability}  \\ \hline 
% C2 Downlink & 40 ms  & 99.9\% \\
% C2 Uplink & 40 ms & 99.9\% \\
% Video Uplink LOS & 1 s & 99.9\% \\
% Video Uplink NLOS & 140 ms & 99.99\% \\ \hline
% \end{tabular}
% \caption{KPIs for command and control and video of \gls{uav} operation}
% \label{tab:qos_uavs}
% \end{table}

%describe the setup with image: Ask Sakthivel to write something about OAI and the BS. 
%messages flow
The office space used in our experiments is illustrated in Fig.~\ref{fig:ExpSetup} and comprises a \gls{5g} Aero \gls{uav} and a \gls{gnb}. The latter is implemented via a \gls{usrp} X410 connected to a host machine (left part of Fig.~\ref{fig:ExpOverview}) where the \gls{5g} protocol stack (both \gls{ran} and core network) runs on the \gls{oai} open-source software. First, the 5G Aero establishes connectivity with the \gls{gnb} via the Sierra Wireless \gls{5g} module, then generates both \gls{c2} and video streaming traffic. Our goal is to measure and evaluate latency and throughput for all packets.
% Our goal is to transmit and analyze the latency and error rate, of \gls{c2} communications and video streaming from the \gls{uav} to the \gls{gnb} using \gls{5g} communication.

The system operates at a 3.6 GHz central frequency and a 40 MHz bandwidth. On the \gls{gnb} side, the \gls{c2} packets are captured by the ground control station application \textit{MissionPlanner}, while the video is captured by a VLC application running in the background. 
In our experiments, we consider four distinct scenarios as described in Fig.~\ref{fig:ExpSetup}. For each scenario, the experiment is executed three times for two minutes. 

The following scenarios are analyzed: In the first two, static flight and \gls{los} conditions are considered, with the 5G Aero hovering 3m and 7m away from the \gls{gnb}, respectively. In the third scenario, we consider static and \gls{nlos} conditions with the 5G Aero 14m away from the \gls{gnb}. Finally, the forth case is a dynamic flight case where the 5G Aero is initially located 3m away from the \gls{gnb} (\gls{los}) and then reaches the last position, i.e., 14m away from the \gls{gnb} in \gls{nlos} conditions.

% 1. Static flight at 3m \gls{los} from the \gls{bs}.
% 2. Static flight at 7m \gls{los} from the \gls{bs}.
% 3. Static flight at 14m \gls{nlos} from the \gls{bs}.
% 4. Dynamic flight varying between 3m and 15m, alternating between \gls{los} and \gls{nlos}.

During each phase, the \gls{uav} transmits both flight and \gls{uav} telemetry data, as well as video streams at different resolutions: 480p, 720p, and 1080p. To comply with regulations, the drone is manually controlled by a pilot who handles takeoff, the \gls{uav} maintains an altitude of 1.5m via lidar-based altitude control until it lands. In the dynamic scenario, the pilot also controls the \gls{uav}'s movements.

\begin{figure}[t!]
    \centering
    \begin{subfigure}[t]{0.15\textwidth} % Same as before
        \includegraphics[width=\textwidth]{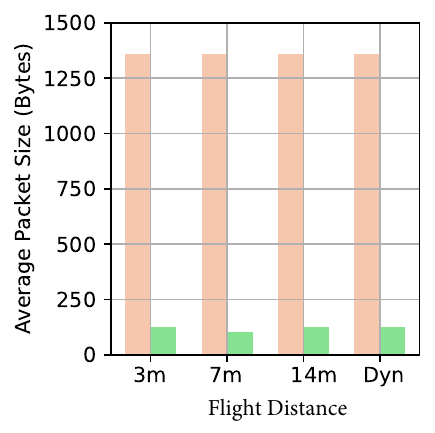}
        \caption{Average Packet size at 1080p resolution.}
        \label{fig:PcktSizeDistance}
    \end{subfigure}
    %\hfill
    \begin{subfigure}[t]{0.15\textwidth} % Same as before
        \includegraphics[width=\textwidth]{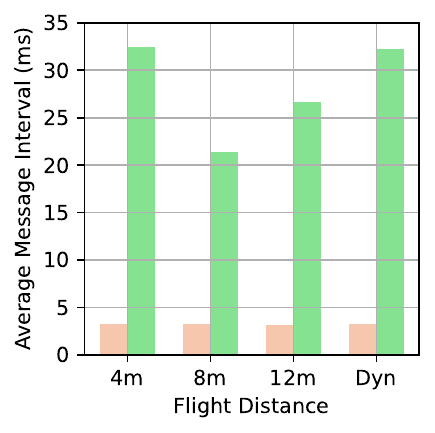}
        \caption{Average Message Interval at 1080p resolution.}
        \label{fig:PcktRateDistance}
    \end{subfigure}
    %\hfill
    \begin{subfigure}[t]{0.15\textwidth} % Same as before
        \includegraphics[width=\textwidth]{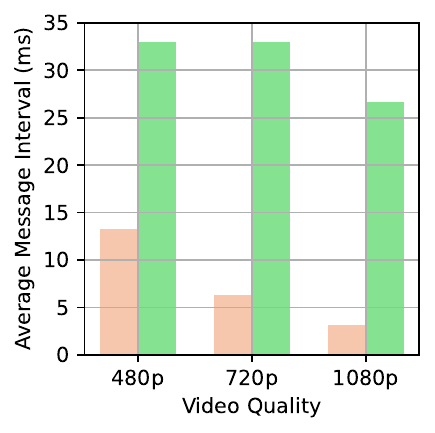}
        \caption{Average Message Interval at 12m distance.}
        \label{fig:PcktRateQuality}
    \end{subfigure}
    % Insert the legend below the subfigures
    \includegraphics[width=0.3\textwidth]{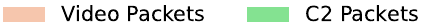} % Adjust path as needed
    \label{fig:Legend}
    \caption{Uplink \gls{c2} and Video Stream performance overview in the four considered scenarios for different distances and video resolutions.}
    \label{fig:pcktSizeOverview}
    \vspace{-0.4cm}
\end{figure}

\section{Results}
\label{sec:results}
To evaluate the performance experienced by the 5G Aero and obtain end-user metrics such as latency and throughput, data collection is performed by capturing packets at both \gls{ue} and \gls{bs} sides using \textit{Wireshark}. 
% describe plots

\subsection{Uplink Traffic Analysis}

Fig.~\ref{fig:PcktSizeDistance} shows that the average packet size in the uplink for both video and \gls{c2} remains consistent at different flight distances, as well as the interval rate at which video packets are transmitted as shown in Fig.~\ref{fig:PcktRateDistance}. Conversely, the interval rate for the \gls{c2} packets varies depending on the flight dynamics. 
% Although the drone remained stationary during all three flights, the minor adjustments and corrections it made to maintain stability resulted in fluctuations in the frequency of \gls{c2} packets transmitted to the ground control station. 
\begin{figure*}[t!]
    \centering
    % First two figures side by side
    \begin{subfigure}[t]{0.32\textwidth}
        \includegraphics[width=\textwidth, height=9cm]{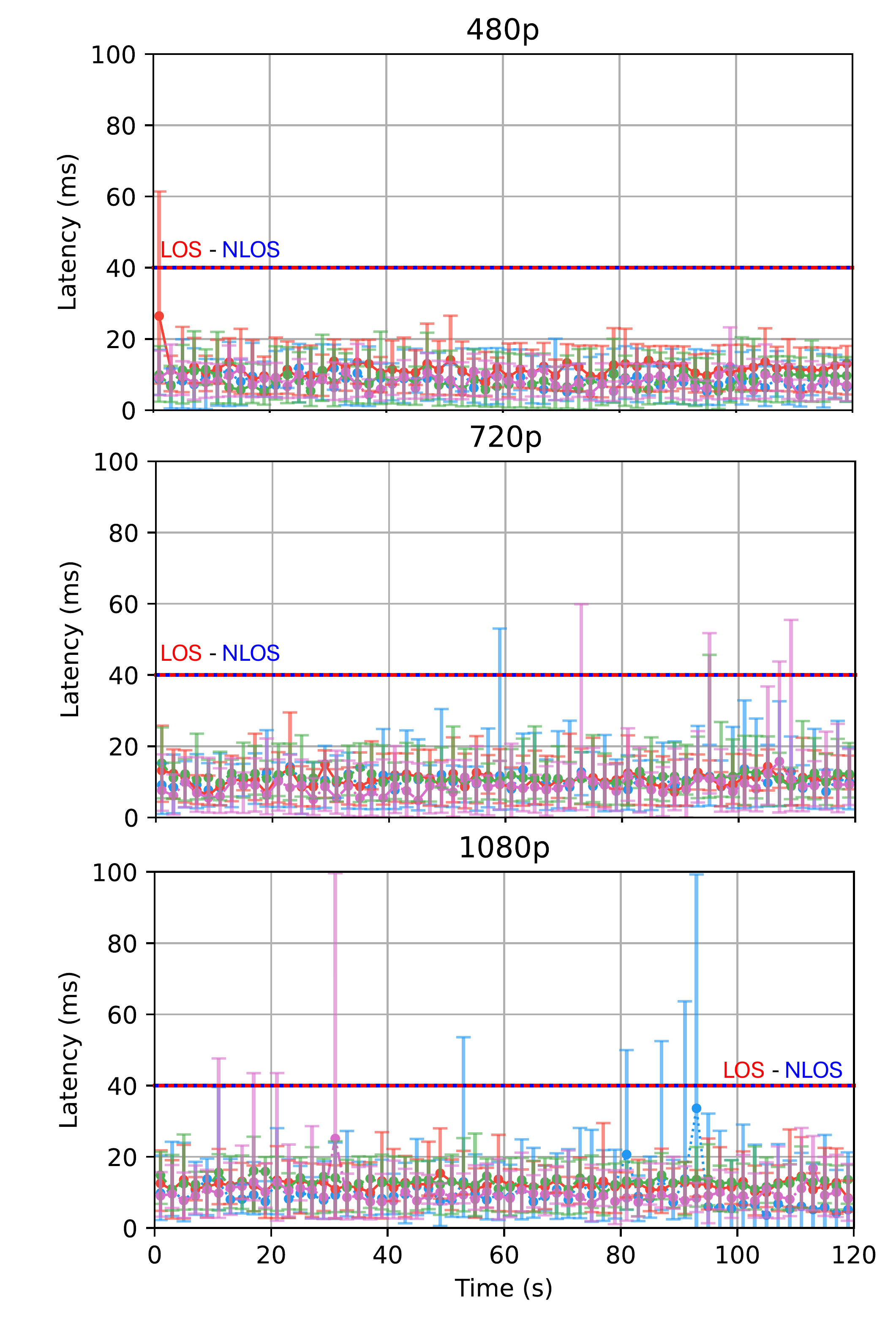} % Adjust the file path as necessary
        \caption{Uplink Latency for \gls{c2} Packets}
        \label{fig:uplink_latency_c2}
    \end{subfigure}
    %\hfill % Adds horizontal space between the figures
    \begin{subfigure}[t]{0.32\textwidth}
        \includegraphics[width=\textwidth, height=9cm]{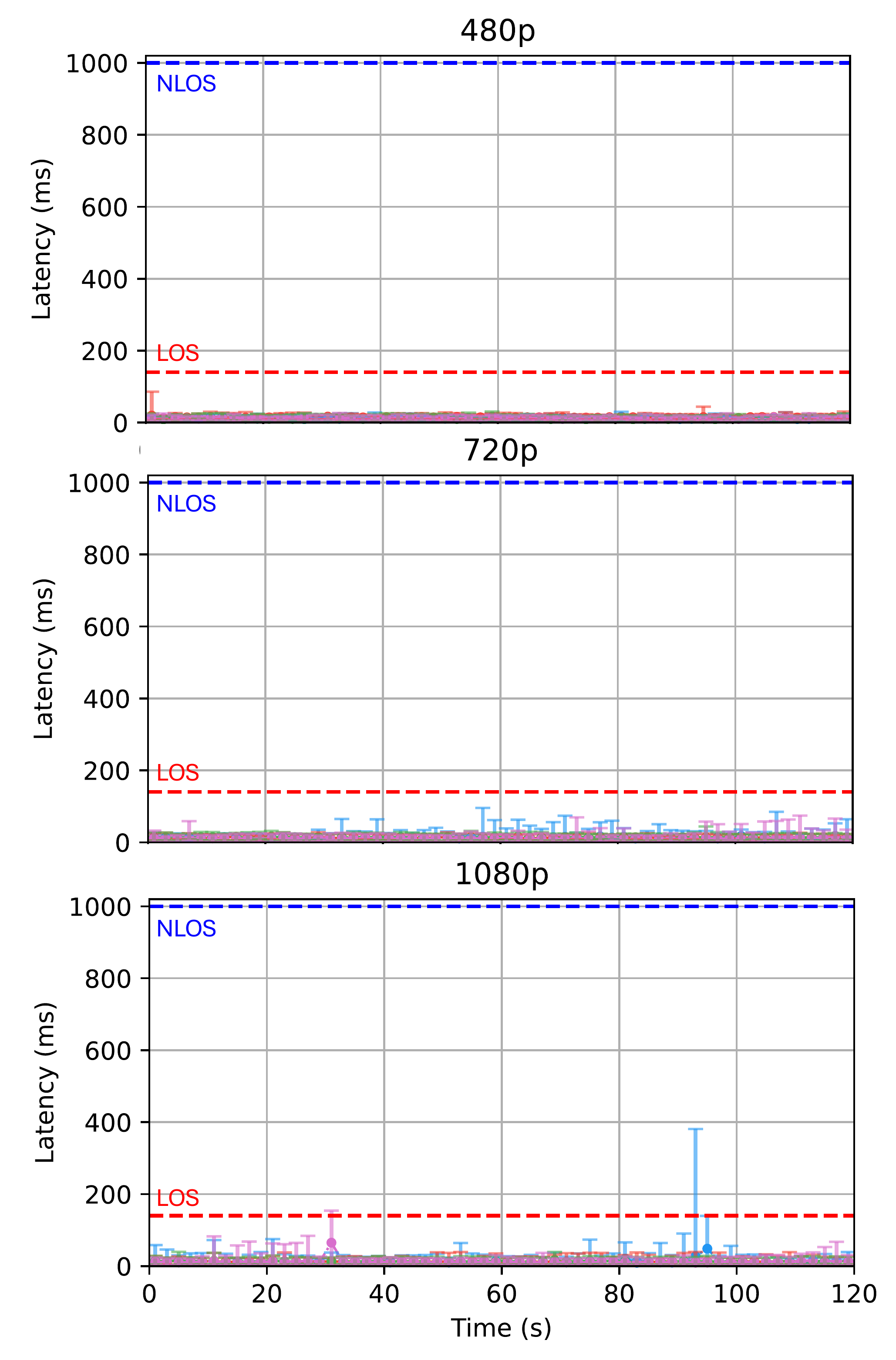} % Adjust the file path as necessary
        \caption{Uplink Latency for Video Packets}
        \label{fig:uplink_latency_video}
    \end{subfigure}
    %\hfill % Adds horizontal space between the figures
\begin{subfigure}[t]{0.32\textwidth}
    \centering
    \includegraphics[width=\textwidth, height=9cm]{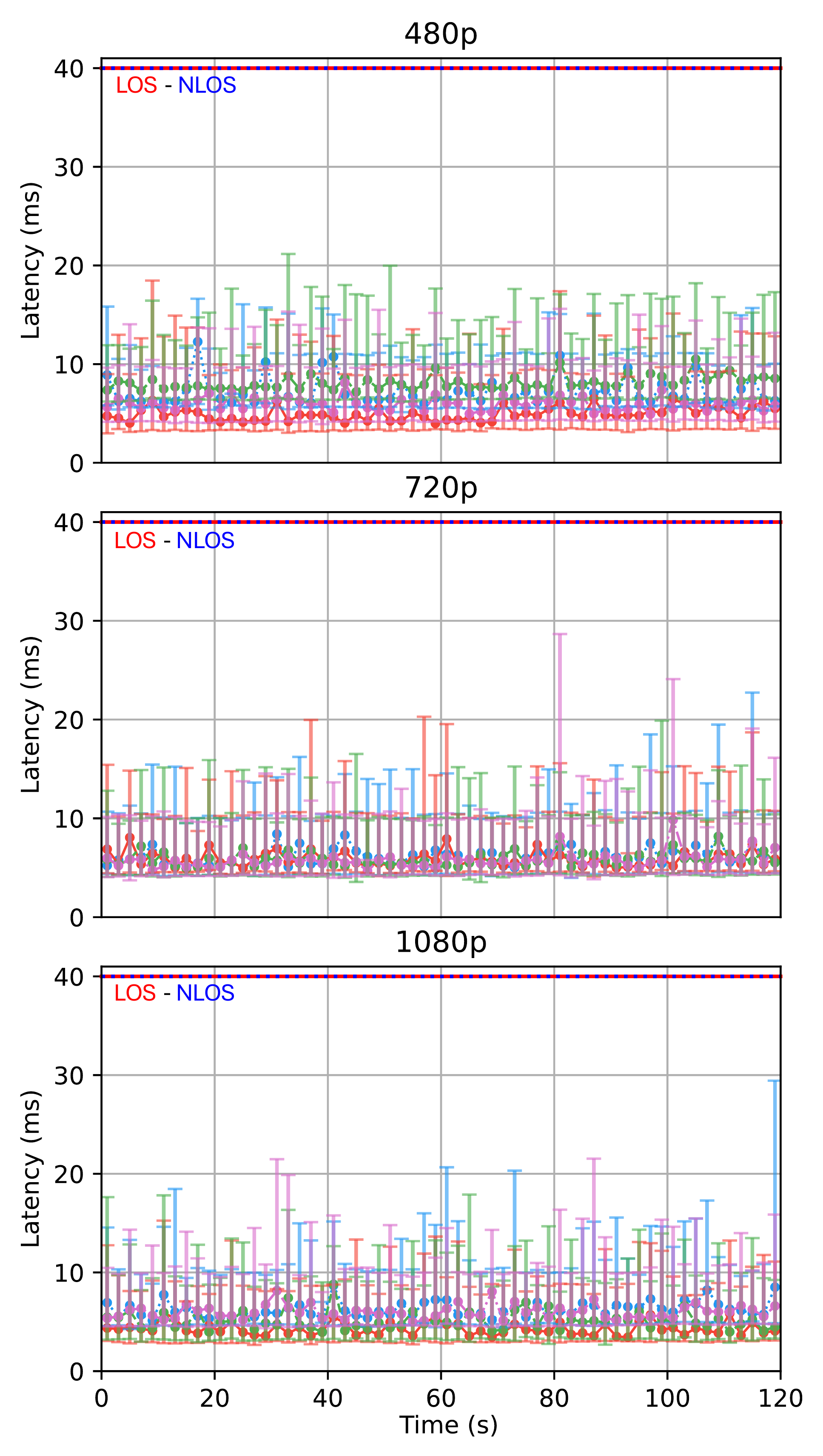} % Force the image height to 5cm
    \caption{Downlink Latency for \gls{c2} Packets}
    \label{fig:downlink_latency_c2}
        \vspace{0.2cm}
\end{subfigure}
    % Third figure spanning full width below the first two
    % \begin{subfigure}[b]{\textwidth}
    %     \centering
    %     \includegraphics[width=0.4\textwidth]{figures/results/RSRPLegend.pdf} % Adjust the file path and width as necessary
    %     %\caption{Legend for Latency Measurements}
    %     \label{fig:latency_legend}
    % \end{subfigure}
    % Insert the legend below the subfigures
    \includegraphics[width=0.4\textwidth]{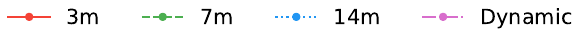}
    \vspace{-0.2cm}
    \caption{Comparison of Uplink and Downlink Latency measurements across different types of packets.}
    \label{fig:latency_comparisons}
    \vspace{-0.4cm}
\end{figure*}
From a different perspective, Fig.~4c illustrates the average packet interval at various video qualities. As previously explained, the frequency rate of \gls{c2} packets fluctuates, while the frequency of video packets changes due to the higher throughput required by the different video qualities. 
The increase in throughput may create a bottleneck in the uplink, hence increasing drastically the latency or the packet error rate.
Specifically, Figs.~\ref{fig:uplink_latency_c2} and \ref{fig:uplink_latency_video} illustrate the uplink latency for \gls{c2} and video packets across different video transmission qualities of 480p, 720p, and 1080p and operational scenarios. In the figure, we also use the dashed horizontal lines to represent the 3GPP-recommended maximum reference values for latency.

\begin{figure*}[t!]
    \centering
    \begin{subfigure}[t]{\textwidth} % Same as before
    \centering
        \includegraphics[width=0.9\textwidth]{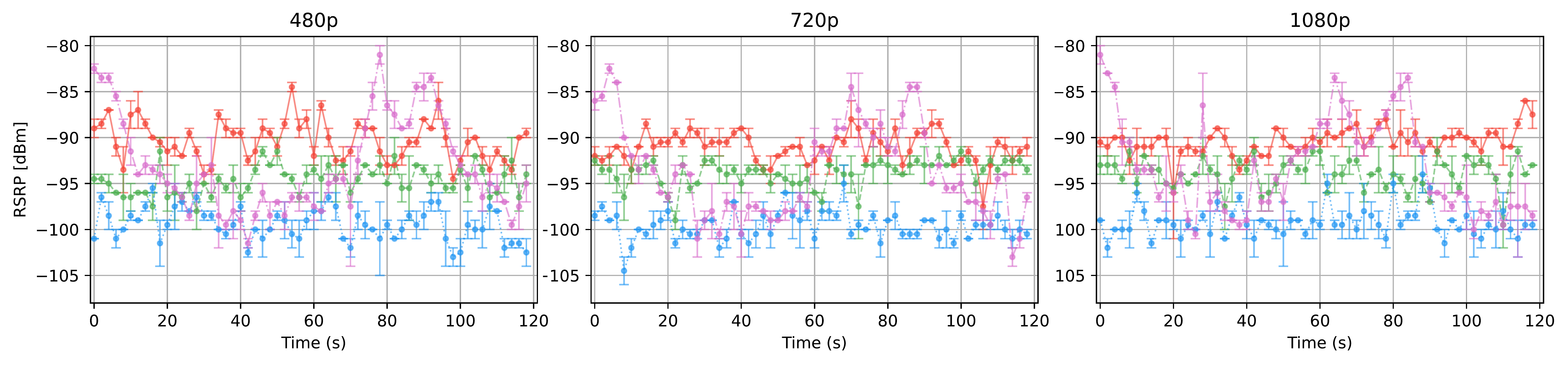}
        \vspace{-0.5cm}
        \label{fig:PcktRateQuality}
    \end{subfigure}
    % Insert the legend below the subfigures
    \includegraphics[width=0.4\textwidth]{figures/results/RSRPLegend.pdf} % Adjust path as needed
    \vspace{-0.2cm}
    \label{fig:Legend}
    \caption{RSRP measured over time for different scenarios and video resolutions.}
    \label{fig:rsrp}
    \vspace{-0.4cm}
\end{figure*}

\subsubsection{480p} 

Fig.~\ref{fig:uplink_latency_c2} (top) shows that  the latency, experienced by the 5G Aero when the resolution is 480p, remains below the desired 40ms threshold for almost the entirety of the experiment. However, we notice that an exception occurs during the takeoff phase when the distance is set to 3m. In this case, we expect a strong reflected path from the ground, which can cause multi-path interference, momentarily increasing latency and resulting in 99.31\% of packets having a latency below the requested threshold. A similar latency spike is observed at takeoff for video packets, as shown in Fig.~\ref{fig:uplink_latency_video} (top). Specifically, we notice that latency reaches up to 830ms, yet this value still falls within the acceptable latency threshold of 140ms defined by 3GPP. Besides the take-off in the 3m experiment, transmission latencies always remain within the specified latency requirements throughout all the \gls{los} and \gls{nlos} experiments.

\subsubsection{720p} 
As we increase the resolution to 720p, the transmission frequency for video packets doubles to 0.006 packets/s, compared to 0.013 packets/s at 480p, potentially increasing congestion on the 5G link. Fig.~\ref{fig:uplink_latency_c2} (center) shows the latency uplink for \gls{c2} packets, revealing notable latency spikes during the 14m and dynamic experiments, both conducted in \gls{nlos} conditions. Specifically, we notice that the 5G Aero in the 14m static experiment achieves a reliability of 99.9\%, with only one packet exceeding a 40ms latency. Conversely, the Dynamic experiment, which alternates between \gls{los} and \gls{nlos}, exhibits a latency reliability of 99.5\%. 
To better understand why latency values increase substantially in \gls{nlos} and dynamic scenarios, Fig.~\ref{fig:rsrp} (center) shows the \gls{rsrp} levels over time between the \gls{uav} and the \gls{gnb}. Throughout the 14m experiment, it is evident that the \gls{rsrp} levels are generally low, with an average value of approximately -100 dBm, largely due to obstacles and absorbing materials. Similarly, in the dynamic experiment, the signal strength fluctuates depending on the drone's position relative to \gls{los} or \gls{nlos}. When in \gls{los}, the signal maintains acceptable levels (approximately -85 dBm); however, as it transitions to a \gls{nlos} scenario, the signal strength diminishes to as low as -100 dBm, resulting in increased delays in packet delivery. Despite this, video packet latencies in Fig.~\ref{fig:uplink_latency_video} remain under 1s, with the worst-case scenario at 0.093s. Finally, transmission latencies for \gls{c2} packets from the 3m and 7m \gls{los} experiments are always under the latency requirements.  

\subsubsection{1080p}

Increasing the video resolution to 1080p results in a transmission rate of 0.003 packets/s, which is double that of 720p and four times that of 480p, further saturating the uplink channel. This increase, combined with low signal quality in \gls{nlos} conditions, which degrades to -103 dBm as shown in Fig.~\ref{fig:rsrp} (right), induces latency spikes. From Fig.~\ref{fig:uplink_latency_c2} (bottom), the 14m experiment in \gls{nlos} conditions, displays a reliability of 98.73\%, and the dynamic flight experiences spikes solely in \gls{nlos} conditions, achieving a reliability of 99.58\%. From Fig.~\ref{fig:uplink_latency_video}, video packets consistently meet the sub-1s latency requirement in both \gls{los} and \gls{nlos}.

\subsection{Downlink Traffic Analysis}

The only data transmitted in downlink are TCP acknowledgment packets related to telemetry traffic. The average packet size is 66.8 bytes, and the \gls{3gpp} latency requirements for this type of downlink traffic are set to 40ms. From Fig.~\ref{fig:downlink_latency_c2}, it is evident that all acknowledgment packets, across all video qualities, are delivered within the required latency, with the highest latency recorded at 28ms. 

\textbf{Battery Consumption:} 
% An essential aspect of UAV performance in different communication environments involves assessing the impact of 5G connectivity on battery life. 
In our tests, the 5G Aero battery, with a capacity of 83Wh, provides an average of 5 minutes and 2 seconds of flight time without the \gls{5g} module turned on. When \gls{5g} transmissions are active, the average flight duration decreases by approximately 1\%, resulting in an average flight time of roughly 4 minutes and 59 seconds. This reduction demonstrates that the \gls{5g} connectivity module in the Sierra Wireless board increases battery drain. However, it represents a relatively negligible decrease if compared to the consumption caused by propellers and flight operations. 

\section{Results Overview and Challenges}
\label{sec:challenges}
Our experiments have shown that the 5G Aero is able to meet the latency requirements set by \gls{3gpp}, with a few exceptions during \gls{nlos} conditions where high throughput demands from videos and obstacles cause interference. It is also noteworthy that the frequency of \gls{c2} packet transmission in both uplink and downlink ranges from 0.026 packets/s to 0.032 packets/s, which exceeds the expectations set by the \gls{3gpp} standards of 0.04 packets/s.

\subsection{Challenges:}
%During the \gls{uav} \gls{5g} Aero's development, we faced several challenges.
During the development, we faced several challenges.

\textbf{i) Size and Configuration:}
The main challenge was designing the drone to fit all components within the compact Intel Aero frame. This required careful placement, disaggregation of the \gls{fcu} and Computing Unit, and the addition of a power converter, while optimizing space and preventing short circuits. Despite the addition of the \gls{5g} module, battery consumption concerns remained minimal, with only a negligible reduction in flight time. Additionally, an alternative approach to further optimize space would be to use an M.2 compute unit capable of directly hosting the Sierra Wireless M.2 5G modem. This would eliminate the need for an external enclosure, reducing weight and further improving the form factor of the \gls{uav}.

\textbf{ii) Electromagnetic Interference}
Upon installing the \gls{5g} radio, we faced issues related to flight stability. The radio's metal casing caused electromagnetic interference, affecting unstable measurements of the IMU functionality placed in the \gls{fcu}. To mitigate this electromagnetic noise, we applied copper foil to the casing.

\section{Conclusions}
\label{sec:conclusions}
This paper has demonstrated the feasibility of utilizing small-factor \glspl{uav} equipped with commercial \gls{5g} radios for a range of 5G-based aerial applications. Our research has led to the development of the compact \gls{5g} Aero \gls{uav}, which delivers \gls{5g} connectivity in line with \gls{3gpp} specifications. While the \gls{5g} Aero \gls{uav} effectively managed diverse experimental scenarios and maintained compliance with stringent latency and reliability standards in all \gls{los} scenarios, it faced a few challenges in \gls{nlos} conditions where high throughput demands and environmental obstacles resulted in poor channel conditions with subsequent increase in latency. Additionally, the impact of \gls{5g} on battery life resulted in a modest reduction of flight duration by approximately 1\%. This slight reduction in battery performance is a small trade-off, considering the substantial benefits brought by \gls{5g} connectivity. In conclusion, this platform not only serves as a robust benchmarking tool for further empirical studies but also establishes a foundational blueprint for future developments in \gls{uav} technology, paving the way for more sophisticated applications and the enhancement of aerial telecommunications infrastructure.

\balance
\bibliographystyle{IEEEtran}
\bibliography{biblio}

\begin{thebibliography}{10}
\providecommand{\url}[1]{#1}
\csname url@samestyle\endcsname
\providecommand{\newblock}{\relax}
\providecommand{\bibinfo}[2]{#2}
\providecommand{\BIBentrySTDinterwordspacing}{\spaceskip=0pt\relax}
\providecommand{\BIBentryALTinterwordstretchfactor}{4}
\providecommand{\BIBentryALTinterwordspacing}{\spaceskip=\fontdimen2\font plus
\BIBentryALTinterwordstretchfactor\fontdimen3\font minus \fontdimen4\font\relax}
\providecommand{\BIBforeignlanguage}[2]{{%
\expandafter\ifx\csname l@#1\endcsname\relax
\typeout{** WARNING: IEEEtran.bst: No hyphenation pattern has been}%
\typeout{** loaded for the language `#1'. Using the pattern for}%
\typeout{** the default language instead.}%
\else
\language=\csname l@#1\endcsname
\fi
#2}}
\providecommand{\BIBdecl}{\relax}
\BIBdecl

\bibitem{agricolture}
P.~Radoglou-Grammatikis, P.~Sarigiannidis, T.~Lagkas, and I.~Moscholios, ``{A compilation of UAV applications for precision agriculture},'' \emph{Computer Networks}, vol. 172, p. 107148, 2020.

\bibitem{dataFromSensor}
A.~V. Sheshashayee, M.~Bordin, P.~B. del Prever, D.~Villa, H.~Cheng, C.~Petrioli, T.~Melodia, and S.~Basagni, ``{Experimental Evaluation of the Performance of UAV-assisted Data Collection for Wake-up Radio-enabled Wireless Networks},'' pp. 01--06, 2024.

\bibitem{aerialBs}
L.~Ferranti, L.~Bonati, S.~D'Oro, and T.~Melodia, ``{SkyCell: A Prototyping Platform for 5G Aerial Base Stations},'' pp. 329--334, 2020.

\bibitem{drivongFromThesky}
M.~Bordin, M.~Giordani, M.~Polese, T.~Melodia, and M.~Zorzi, ``Autonomous driving from the sky: Design and end-to-end performance evaluation,'' in \emph{2022 IEEE Globecom Workshops (GC Wkshps)}, 2022, pp. 1610--1615.

\bibitem{lowLatency}
A.~Masaracchia, Y.~Li, K.~K. Nguyen, C.~Yin, S.~R. Khosravirad, D.~B.~D. Costa, and T.~Q. Duong, ``{UAV-Enabled Ultra-Reliable Low-Latency Communications for 6G: A Comprehensive Survey},'' \emph{IEEE Access}, vol.~9, pp. 137\,338--137\,352, 2021.

\bibitem{ETSI181}
E.~3GPP, ``Unmanned aerial system (uas) support in 3gpp (3gpp ts 22.125 version 18.1.0 release 18),'' May 2024.

\bibitem{ETSI182}
------, ``Support of uncrewed aerial systems (uas) connectivity, identification and tracking; stage 2 (3gpp ts 23.256 version 18.2.0 release 18),'' May 2024.

\bibitem{Buczek_2021}
J.~Buczek, L.~Bertizzolo, S.~Basagni, and T.~Melodia, ``What is a wireless uav?: A design blueprint for 6g flying wireless nodes,'' in \emph{Proceedings of the 15th ACM Workshop on Wireless Network Testbeds, Experimental evaluation and Characterization}, ser. ACM MobiCom ’21, 2021.

\bibitem{airlink}
\BIBentryALTinterwordspacing
{Sky Drones}. (2024) {AIRLink: The Most Advanced AI Drone Flight Controller}. [Online]. Available: \url{https://sky-drones.com/airlink}
\BIBentrySTDinterwordspacing

\bibitem{ferranti2020skycell}
L.~Ferranti, L.~Bonati, S.~D'Oro, and T.~Melodia, ``{{SkyCell}: A Prototyping Platform for {5G} Aerial Base Stations},'' in \emph{Proceeding of IEEE SwarmNet}, Cork, Ireland, August 2020.

\bibitem{nanaoUAV}
P.~Laclau, V.~Tempez, F.~Ruffier, E.~Natalizio, and J.-B. Mouret, ``Signal-based self-organization of a chain of uavs for subterranean exploration,'' \emph{Frontiers in Robotics and AI}, vol.~8, 2021.

\bibitem{1mDrone}
D.~Mishra, H.~Gupta, and E.~Natalizio, ``Sky5g: Prototyping 5g aerial base station (uav-bs) for on-demand connectivity from sky,'' in \emph{2024 IEEE Wireless Communications and Networking Conference (WCNC)}, 2024, pp. 1--6.

\bibitem{oai}
F.~Kaltenberger, T.~Melodia, I.~Ghauri, M.~Polese, R.~Knopp, T.~T. Nguyen, S.~Velumani, D.~Villa, L.~Bonati, R.~Schmidt, S.~Arora, M.~Irazabal, and N.~Nikaein, ``Driving innovation in 6g wireless technologies: The openairinterface approach,'' 2025.

\end{thebibliography}

\end{document}